\documentclass[journal]{IEEEtran}
\usepackage{cite}

\ifCLASSINFOpdf
  \usepackage[pdftex]{graphicx}
  \graphicspath{{./Images/Simu/}{./Images/Exp/}{./Images/Exp/Autoradar_new/}}
\else
  \usepackage[dvips]{graphicx}
\fi
\usepackage{color}

\usepackage{multirow}
\usepackage{threeparttable}

\usepackage{amsmath}
\usepackage[ruled]{algorithm2e}

\ifCLASSOPTIONcompsoc
  \usepackage[caption=false,font=normalsize,labelfont=sf,textfont=sf]{subfig}
\else
  \usepackage[caption=false,font=footnotesize]{subfig}
\fi

\usepackage{stfloats}
\begin{document}
%
\title{CFAR-Based Interference Mitigation for FMCW Automotive Radar Systems}
%
%
%

\author{Jianping~Wang,~\IEEEmembership{Member, IEEE}
\thanks{The author is with the Faculty of Electrical Engineering, Mathematics and Computer Science (EEMCS), Delft University of Technology, Delft, 2628CD, the Netherlands e-mail: J.Wang-4@tudelft.nl.
}
}

%
%

\markboth{Journal of \LaTeX\ Class Files,~Vol.~, No.~, }
{Shell \MakeLowercase{\textit{et al.}}: Bare Demo of IEEEtran.cls for IEEE Journals}
%



\vspace{-3mm}
\maketitle

\begin{abstract}	
In this paper, constant false alarm rate (CFAR) detector-based approaches are proposed for interference mitigation of Frequency modulated continuous wave (FMCW) radars. The proposed methods exploit the fact that after dechirping and low-pass filtering operations the targets' beat signals of FMCW radars are composed of exponential sinusoidal components while interferences exhibit short chirp waves within a sweep. The spectra of interferences in the time-frequency ($t$-$f$) domain are detected by employing a 1-D CFAR detector along each frequency bin and then the detected map is dilated as a mask for interference suppression. They are applicable to the scenarios in the presence of multiple interferences. Compared to the existing methods, the proposed methods reduce the power loss of useful signals and are very computationally efficient. Their interference mitigation performances are demonstrated through both numerical simulations and experimental results.    
\end{abstract}


%
\IEEEpeerreviewmaketitle

\section{Introduction}
%
%
%
%
\IEEEPARstart{N}{owadays} frequency modulated continuous wave (FMCW) radars have become a key device for automotive assistant/autonomous driving due to its operational capability in all day time and all weather conditions as well as its low cost. With the increase of vehicles equipped with radar sensors, the FMCW radar systems mounted on different cars in busy area will inevitably suffer from strong interfering influence from the radar systems on the neighboring cars as well as other radars on the same car when they operate at the same time. The strong interferences would cause significantly increased noise floor, weak target mask and reduced probability of target detection. Therefore, to overcome these risks, effectively mitigating interferences from other radars is critical to high-performance automotive radars.

Interference mitigation (IM) for automotive radar is a hot topic in recent years. In the literature, many approaches have been proposed and developed to suppress the interferences among different automotive radars, which can be classified into three categories: radar system coordination, radar system design and waveform design, and signal processing. For the radar system coordination approaches, a coordination scheme, which is either centralized \cite{Khoury2016} or distributed\cite{Ishikawa2019RADAR,Aydogdu2019TITS}, among different operational radars are devised to avoid conflicts by adjusting the operating parameters (i.e.,transmitting time, spectrum, etc.) of each radar within the interfering area. Although these coordination schemes originated from communication network could effectively avoid certain interferences, they usually require to introduce an extra coordination unit to the the existing FMCW radar systems or need communication with a coordination center for a local distributed radar network.

On the other hand, some new radar system architectures and waveforms are proposed to benefit the interference mitigation \cite{Luo2013TCAS,Hu2019TVT,Kitsukawa2019EuRAD,Gambi2008TVT,Xu2018GRSL,Uysal2020TVT,Kim2018IET}. The frequency-hopping random chirp (FHRC) FMCW technique \cite{Luo2013TCAS,Hu2019TVT} and FMCW radar with random repetition interval \cite{Kitsukawa2019EuRAD} resets the parameters of the chirp signals (the bandwidth, sweep duration, center frequency, repetition interval) every cycle to result in noise-like frequency responses of mutual interferences after the received signals are down-converted and demodulated. Both techniques would mitigate partial interferences and avoid the appearance of ghost targets caused by mutual interferences. However, the randomized repetition intervals would cause the Fast Fourier transform, which is conventionally used, inapplicable for the fast Doppler processing. On the other hand, pseudo-random noise signals \cite{Xu2018GRSL} and chaotic sequences \cite{Gambi2008TVT} are proposed to mitigate mutual interferences for automotive radars. For these radar systems, the received signals are processed by the correlation operation and a high sampling frequency is generally required for the Analog-to-Digital Converter (ADC), which would increase the cost of the radar systems. To exploit the advantages of both noise-like signals and the FMCW radar system, phase modulated (PM) FMCW radar systems modulate the FMCW waveforms with orthogonal or random sequences as transmitted signals \cite{Uysal2020TVT,Kim2018IET}. In reception, the received PM-FMCW signals can be down-converted as the traditional FMCW radars and then decoded by correlation with the stored sequences used for transmission modulation. The scattered signals resulting from the transmitted signals generally result in high correlation peaks while the uncorrelated interferences would spread out and build up the noise floor after decoding. Consequently, the raised noise floor could overwhelm the weak targets and reduce the probability of detection. In addition, PM-FMCW radar requires to design a new radar architecture, which cannot be easily implemented with the existing FMCW radar chips.

Moreover, for the FMCW radars, a number of signal processing approaches to interference mitigation have been presented, which includes both traditional signal processing methods and deep-learning based methods. The traditional signal processing methods usually address the interference mitigation by filtering or separating the interferences from the received signals in various domains (i.e., space, time, frequency, time-frequency, etc). For array-based radar system, interference mitigation can be achieved by constructing nulls in the directions of arrival (DOA) of the interferences through beamforming \cite{Bechter2017TMTT,Artyukhin2019EnT,Rameez2018RadarConf}. However, these approaches would suppress targets' signals scattered from the same DOAs of the interferences. In \cite{Nozawa2017radar}, the interference is detected based on a threshold and then suppressed by windowing in time. In \cite{Wu2020IETRSN}, an iterative modified method based on empirical mode decomposition is proposed to decompose the low-pass filter output of an FMCW radar as a series of empirical modes in the time domain while in \cite{Lee2019TITS} the wavelet denoising method is used to separate interferences from the useful signals. Both approaches implicitly assume interferences are sparse in time in the received signals and their performances would degrade with the increase of the proportion of interference-contaminated samples in the acquired signal. By contrast, the Adaptive Noise Canceller (ANC) \cite{Jin2019TVT} is utilized to suppress interferences in the frequency domain. Although it is computationally very efficient, its performance heavily depends on if a proper correlated reference input of the adaptive filter can be found. Meanwhile, in \cite{Neemat2019TMTT} the interference-contaminated signal samples of FMCW radars are first cut out in the short-time-Fourier-transform (STFT) domain and then a Burg-based methods is developed to reconstruct the signal in the cut-out region based on an auto-regressive (AR) model along each frequency bin. However, with the increase of the cut-out region in the signal, the accuracy of the recovered signals with this approach drops rapidly. Moreover, recently some deep-learning approaches are used for interference mitigation of FMCW radars \cite{Mun2020ICASSP,Rock2020RADAR}. These approaches generally require a large volume of dataset acquired in various situations for training.             

In this paper, we proposed two constant false alarm rate detector (CFAR) \cite{Mark2014Book} based approaches to mitigate interferences for FMCW radars. In both approaches, the acquired beat signal is transformed into the time-frequency ($t$-$f$) domain by using the STFT. Then a one-dimensional (1-D) CFAR detector is utilized to detect interferences and the detection map is dilated to generate a mask for interference suppression. Specifically, one approach is to zero out the interference-contaminated samples and the other one is to keep their phases unchanged but correct their amplitudes by the mean of the amplitudes of the interference-free samples in the corresponding frequency bin based on the dilated detection map, which are termed as the CFAR-Zeroing (CFAR-Z) and CFAR-Amplitude Correction (CFAR-AC) approaches in the paper. Compared to the existing approaches, the proposed approaches are capable to mitigate multiple interferences and minimize the power loss of useful signals. Their interference mitigation performance have been validated through both numerical simulations and experimental results. Moreover, they is very efficient and can be implemented for real-time interference mitigation of FMCW automotive radars.  

The rest of the paper is organized as follows. Section~\ref{sec: sig_model_CFAR_IM_method} briefly describes the signal model of the FMCW radar. Then, the CFAR-based interference mitigation approaches are presented in section~\ref{sec:CFAR_based_IM_approaches}.To demonstrate the interference mitigation performance of the proposed approach, numerical simulations and experimental results are shown in sections \ref{sec:numerical_simu} and \ref{sec:exp_result}. Finally, some conclusions are drawn in section~\ref{sec:conclusion}.

\section{Signal model and CFAR-based interference mitigation method} \label{sec: sig_model_CFAR_IM_method}


Assume that the transmitted signal $p(t)$ by an FMCW radar is given by 
\begin{equation}
p(t) = \exp\left[j2\pi \left(f_0 t + \frac{K}{2} t^2 \right) \right]
\end{equation} 
where $f_0$ is the starting frequency of the FMCW sweep, and $K$ is the sweep slope. Considering the single bounce scattering, then the signals scattered back from point-like targets are the superposition of the time-delayed transmitted signals. Meanwhile, assume that the scattered signals from targets are contaminated by an interference $s_\text{int}(t)$ during its reception. After dechirping and low-pass filtering operating on receiver, the acquired beat signals is represented as 
\begin{align} \label{eq:beat_sig_timeDom}
s(t) &= s_b(t) + \tilde{s}_\text{int}(t) + n(t) \nonumber\\
&= \sum_{i=1}^{M} a_i \exp\left(-j2\pi f_{b,i} t\right)  + \mathcal{F}_{lp} \left(s_\text{int}(t)\cdot p^\ast(t) \right) + n(t)
\end{align}
where $\mathcal{F}_{lp}$ is the low-pass filtering operator whose cut-off frequency is determined by the desired maximum detectable range of targets. $s_b(t)=\sum_{i=1}^{M}a_i \exp\left(-j2\pi f_{b,i} t \right) $ is the beat signals of $M$ scatterers, which is composed of $M$ complex exponentials with the beat frequency $f_{b,i}$ and scattering coefficient $a_i$ for the $i^\text{th}$ scatterer. Note here a residual video phase term is subsumed by $a_i$ for conciseness. $\tilde{s}_\text{int}(t) = \mathcal{F}_{lp}(s_\text{int}(t) \cdot p^\ast(t) )$ is the remaining interference after the low-pass filtering, and $n(t)$ denotes the noise and measurement errors. According to the analysis in \cite{Wang2020MP}, the interference $\tilde{s}_\text{int}(t)$ in \eqref{eq:beat_sig_timeDom} generally exhibits as some short chirp-like pulses in the time domain. Although FMCW interferences with the same sweep slope and frequencies falling into the receiving bandwidth would result in ghost targets, its probability is extremely small \cite{Kim2018TVT}. Therefore, After taking the STFT of $s(t)$, the time-frequency ($t$-$f$) domain counterparts of the beat signals of scatterers show as straight lines along the corresponding frequency bins while interferences display as oblique lines, as illustrated in Fig.~\ref{fig:Simu_pointTarget_InterfMitig}\subref{fig:pt_TF_15dB}. These different distributions of useful beat signals and interferences motive us to proposed the CFAR-based interference mitigation approach in the following.

\section{CFAR-based interference mitigation approach}  \label{sec:CFAR_based_IM_approaches}

\begin{algorithm}[!t]
	\SetAlgoLined
	\KwData{Complex signal $\mathbf{s}$ in a sweep}
	\KwResult{Complex signal $\mathbf{s}_c$ after interference mitigation}
	\Begin{
		$\mathbf{S}_{tf} = \textbf{STFT}(\mathbf{s})$;
		$[N_r, N_c] = \textbf{size}(\mathbf{S}_{tf})$;  \\
		$\mathbf{P}_{tf} = \mathbf{S}_{tf} \odot \mathbf{\bar{S}}_{tf} $;  \\
		\For{$k=1$ \KwTo $N_r$}{
			$\mathbf{D}(k,:) =$ \textbf{CFARDetector}$[\mathbf{P}_{tf}(k,:)]$;
		}
		$\mathbf{D}_{dl} =$ \textbf{maskDilate}$(\mathbf{D})$;  \\
		$\mathbf{S}_{tf}(\mathbf{D}_{dl})=0$;    \\
		$\mathbf{s}_c = \textbf{ISTFT}(\mathbf{S}_{tf})$;
	}
	\caption{CFAR-based interference mitigation method.}
	\label{alg:CFAR_based_IM}
\end{algorithm}

Accurately detecting interferences is crucial for effective interference mitigation. Based on the above analysis of different distribution features of useful signals and interferences in the $t$-$f$ domain, i.e., straight lines for useful signals along the frequency bin and oblique lines for interferences, detecting interferences can be converted to distinguish the signals distributed along oblique lines relative to the frequency axis. Therefore, we propose to utilize a 1-D CFAR detector along each frequency bin in the $t$-$f$ domain to detect interferences and then suppress them.

The complete CFAR-based interference mitigation method is shown in Algorithm~\ref{alg:CFAR_based_IM}. In principle, it contains three major steps in implementation, which are described in detail as follows.

\subsection{Time-Frequency Analysis with the STFT}
Applying the STFT to the acquired signal in \eqref{eq:beat_sig_timeDom}, its $t$-$f$ spectrum is obtained as
\begin{equation}
S(\tau,f) = \int_{-\infty}^{\infty} s(t) w(t-\tau) e^{-j2\pi f t} dt
\end{equation}
where $w(\tau)$ is the window function, for instance, a Gaussian window or Hann window. For $N$ discrete signal samples $s[k]=s(k\Delta t)$, $k=0,1,\cdots,N-1$, the discrete $t$-$f$ spectrum samples over a regular grid are generally computed by 
\begin{equation} \label{eq:discrete_STFT_signal}
\begin{aligned}
S_{tf}[m,n] &= S(m\Delta \tau, n\Delta f) \\
&=\sum_{k=0}^{N-1} s(k\Delta t) w(k\Delta t-m\Delta \tau) e^{-j2\pi nk\Delta f \Delta t} \Delta t
\end{aligned}
\end{equation}
where $\Delta t$ is the time sampling interval, $\Delta\tau$ is the sliding step of the window and $\Delta f$ is the step of frequency samples. One can see that for a fixed time delay $m\Delta \tau$ of the window, \eqref{eq:discrete_STFT_signal} can be efficiently implemented by using the fast Fourier transform (FFT). For the convenience of computation, generally $\Delta \tau=l\cdot\Delta t$ and $l\geq 1$ is an integer. Sliding the window over the signal duration, the $t$-$f$ spectrum is obtained as a two-dimensional matrix with dimensions of $N_t\times N_f$ along the time and frequency axes, respectively, where $N_t$ is the number of sliding steps of the time window and $N_f$ is the number of FFT points.  

Then, the spectrogram is obtained as the amplitude squared of the $t$-$f$ spectrum, given by
\begin{equation}
P_{tf}[m,n] = \left|S_{tf}[m,n]\right|^2 = S_{tf}[m,n] \cdot \bar{S}_{tf}[m,n]
\end{equation}
where $\bar{S}_{tf}$ is the complex conjugate of $S_{tf}$.

\subsection{CFAR Detection and Detection Mask Dilation}
In this step, the interference detection is performed. After getting the power spectrogram, a Cell Averaging CFAR (CA-CFAR) detector \cite{Mark2014Book} is utilized to the spectrum density along each frequency bin, resulting in a detection matrix $\mathbf{D}$ with the same size as the spectrogram. The detection matrix $\mathbf{D}$ has the entries of ones and zeros and the entries of one indicate the positions of the detected interferences. The numbers of guard cells and training cells, the probability of false alarm and the threshold factor of the CFAR detector can be set based on the different scenarios. 

After acquiring the detection map with the CFAR detector, in principle it could be employed as a mask to suppress interferences. However, due to the possible existence of several interference-contaminated spectral samples in a frequency bin, a relatively large threshold value would be calculated; thus, it causes the missed detection of some edge cells of the interferences. To alleviate such problem, a dilation procedure \cite{Gonzalez2009}, which is widely used for image processing, is introduced to slightly swell the detected mask of interferences. Considering the detection map $\mathbf{D}$ as a binary image, the one-valued pixels form a pattern of the detected interferences, denoted as $I$. To dilate the pattern $I$, a \textit{structuring element} $B$ is used and its origin is translated throughout the entire domain of the input image $\mathbf{D}$. The dilation of the pattern $I$ by the structuring element $B$ is defined as the set operation
\begin{equation}
I_{dl} = I\oplus B=\left\{z|(\hat{B})_z \cap I \neq \emptyset \right\}
\end{equation}
where $\hat{B}$ is the reflection of  the structuring element $B$ about its origin and $z$ indicates the location that the origin of the structuring element is translated to. So the dilated pattern $I_{dl}$ is the set of pixel locations $z$, where the reflected structuring element overlaps with at least one element in $I$ when translated to $z$. Accordingly, at these locations of $z$, the output image $\mathbf{D}_{dl}$ is 1, which contains the dilated pattern $I_{dl}$. Since the detected pattern of interferences is some oblique thick lines with possible round ends, the disk-shaped or octagonal structuring element can be used. 

\subsection{Interference Mitigation and Signal Recovery}
The dilated detection map of interferences can be used as a mask for interference mitigation. With the aid of the dilated detection map, a simplest interference mitigation approach is to zero out the interference-contaminated signal samples in the $t$-$f$ spectrum $S_{tf}$, denoted as CFAR-Z for conciseness in the following. However, the zeroing operation suppresses not only interferences but also the useful signals, thus causing the power loss of the targets' signals.

To circumvent the signal power loss of the CFAR-Z method, we suggest utilizing the amplitude correction method \cite{Wagner2018ISCAS} to the interference-contaminated samples based on the CFAR detection map. The resultant approach is termed as CFAR-AC. The basic idea of this approach is to replace the amplitudes of the interference-contaminated samples with the average amplitude of the interference-free spectrum samples in the corresponding frequency bin but keep their phases invariant. The new value for a interference-contaminated sample $S_{tf}[m_i,n_i]$ is given by
\begin{equation}
\tilde{S}_{tf}[m_i,n_i] = A_{n_i} e^{j\arg(S_{tf}[m_i,n_i])}
\end{equation}
where $\tilde{S}_{tf}[m_i,n_i]$ is the new sample value at the position $[m_i,n_i]$ obtained after interference mitigation, and $\arg(x)$ takes the phase of a complex number $x$. $A_{n_i}$ is the average amplitude of the interference-free samples in the ${n_i}^\text{th}$ frequency bin. In this way, the strong power of the interferences is significantly suppressed. 

After that, an inverse STFT (ISTFT) is applied to the interference-mitigated $t$-$f$ spectrum to recover the targets' beat signals in the time domain.

In addition, we should mention that although the phases of the new sample values still surfer from the disturbance of interferences, their effects are negligible after taking further coherent range compression and/or Doppler processing. Moreover, for the array signals contaminated simultaneously by the same interferences, CFAR-AC approach has no impact on the beamforming performance as the phases of signals are kept unchanged.

\section{Numerical simulations} \label{sec:numerical_simu}

{
	\renewcommand{\arraystretch}{1.2}
	\begin{table}[!t]
		\centering
		\caption{Parameters for numerical simulations}
		\begin{tabular}{l|l}
			\hline\hline 
			\textbf{Parameter}    & \textbf{Value}  \\ \hline
			Center frequency    & $77\,\mathrm{GHz}$  \\ \hline
			Bandwidth & $600\,\mathrm{MHz}$ \\ \hline
			Sweep duration of FMCW signal & $100\,\mu \mathrm{s}$ \\ \hline
			Maximum detection range & $250\,\mathrm{m}$ \\ \hline
			Sampling frequency & $40\,\mathrm{MHz}$ \\  \hline
			Distances of three point targets & $30$, $80$, and $150\,\mathrm{m}$\\ \hline
			\hline 
		\end{tabular}
		\label{tab:simu_parameters}
	\end{table}
}

\begin{figure*}[!t]
	\centering
	\vspace{-3mm}
	\subfloat[]{
		\includegraphics[width=0.325\textwidth]{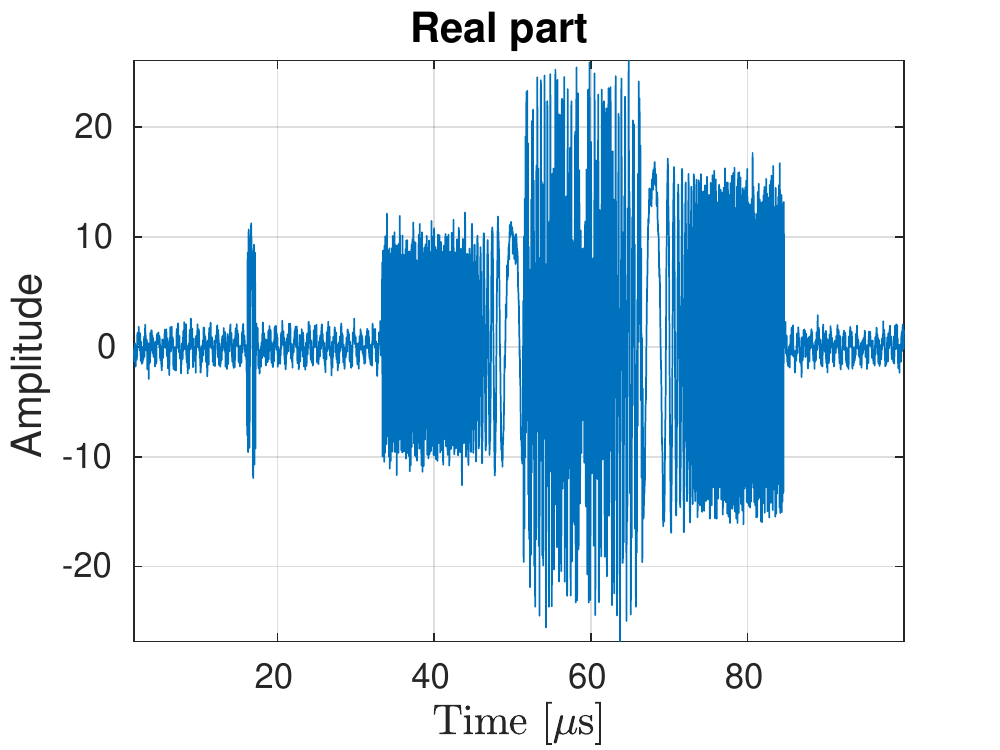}
		\label{fig:pt_RawSig_RealPart_15dB}
	}
	\subfloat[]{
		\includegraphics[width=0.325\textwidth]{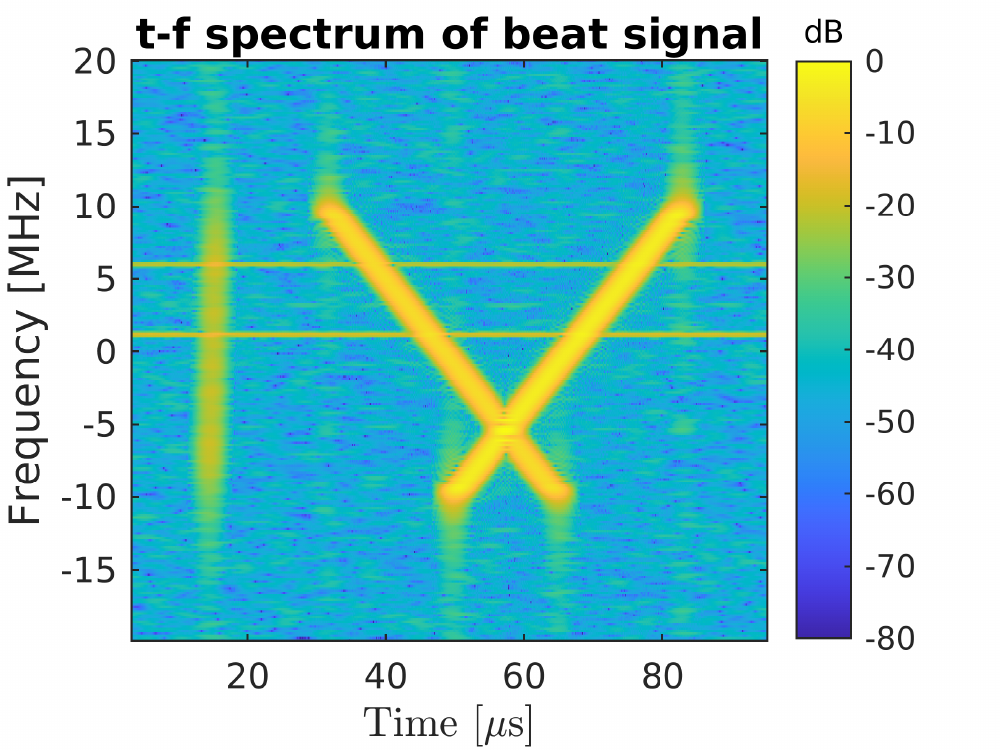}
		\label{fig:pt_TF_15dB}
	}
	\subfloat[]{
		\includegraphics[width=0.325\textwidth]{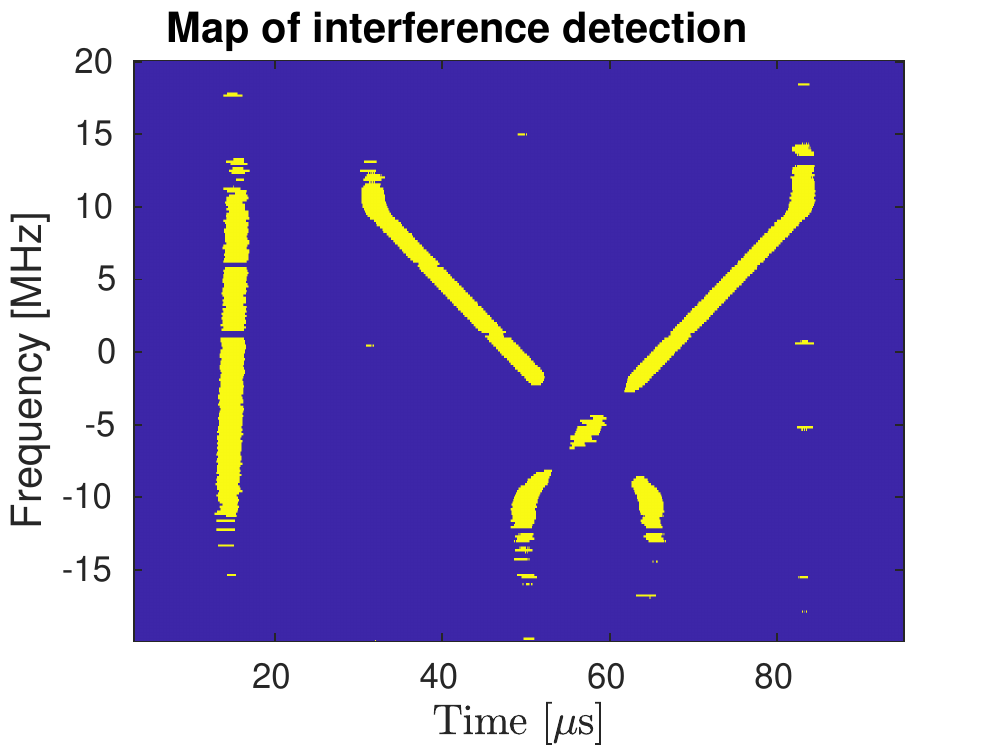}
		\label{fig:pt_DetMAP_15dB}
	}
	
	\vspace{-1mm}
	\subfloat[]{
		\includegraphics[width = 0.325\textwidth]{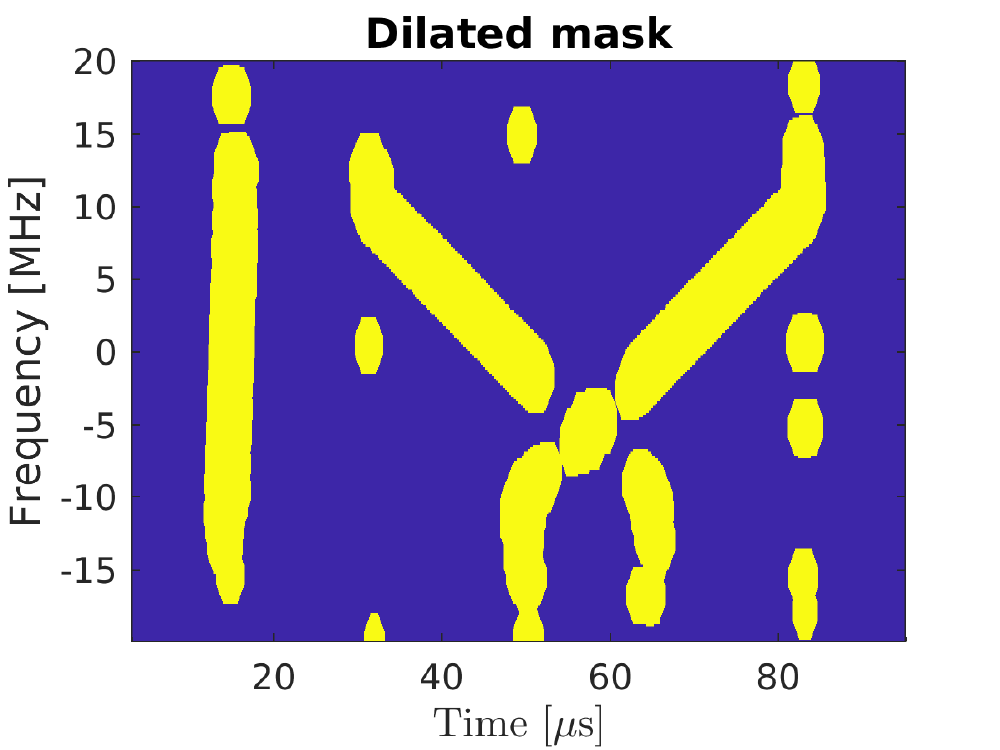}
		\label{fig:pt_CFAR_mask_15dB}
	}
	\subfloat[]{
		\includegraphics[width=0.325\textwidth]{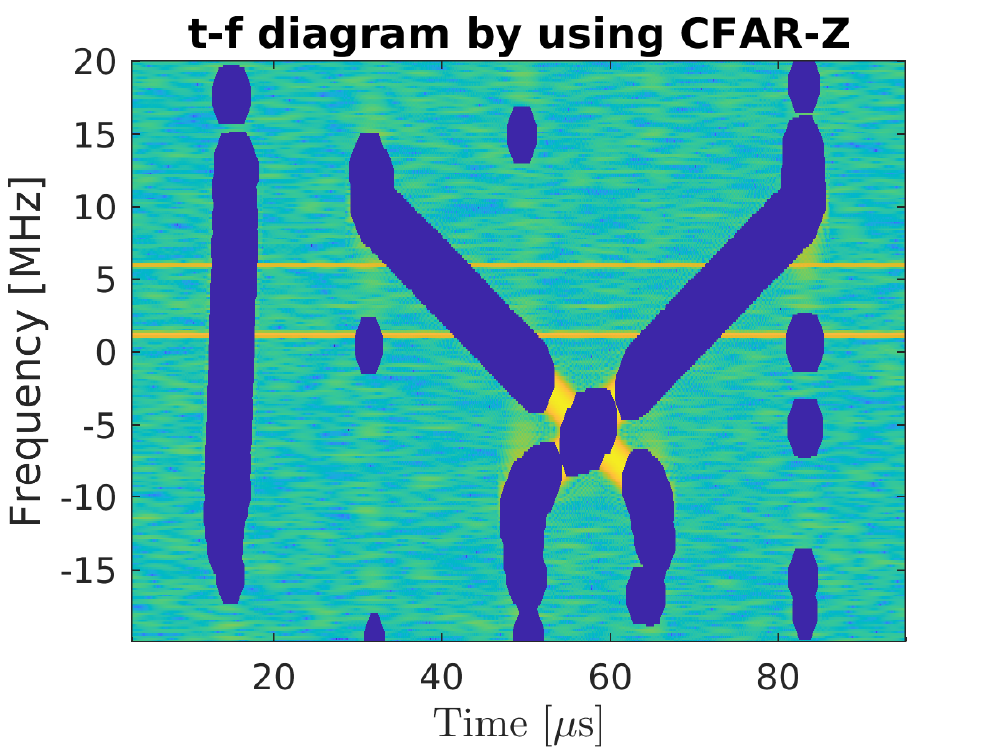}
		\label{fig:pt_TF_15dB_cfarZ}
	}
	\subfloat[]{
		\includegraphics[width=0.325\textwidth]{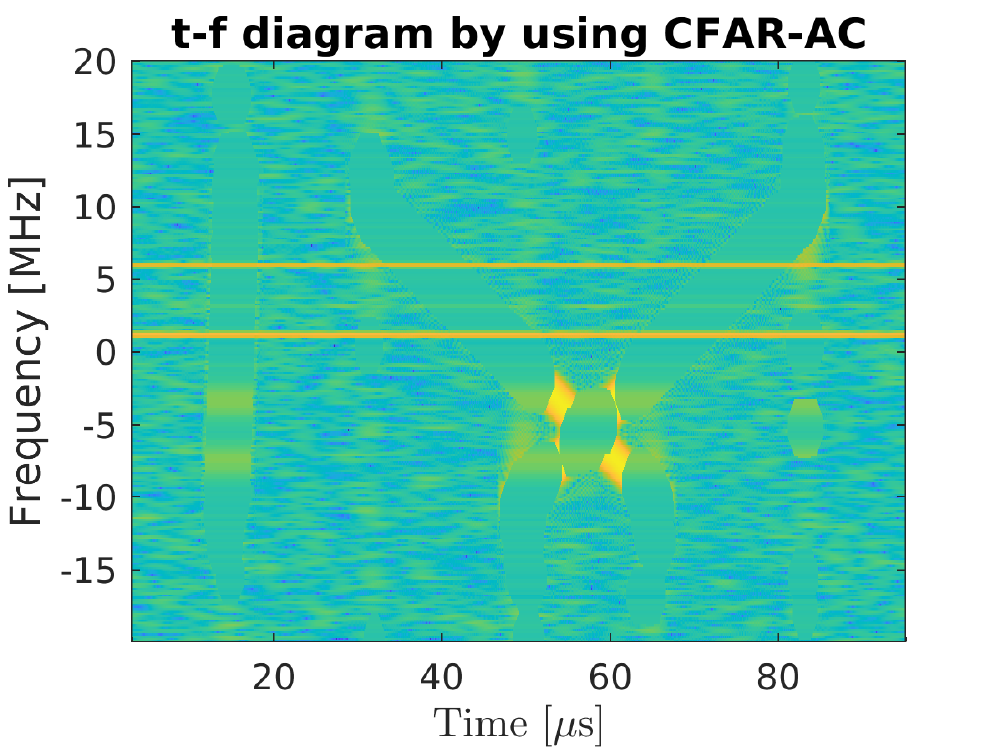}
		\label{fig:pt_TF_15dB_cfarAC}
	}
	
	\vspace{-1mm}
	\subfloat[]{
		\includegraphics[width=0.325\textwidth]{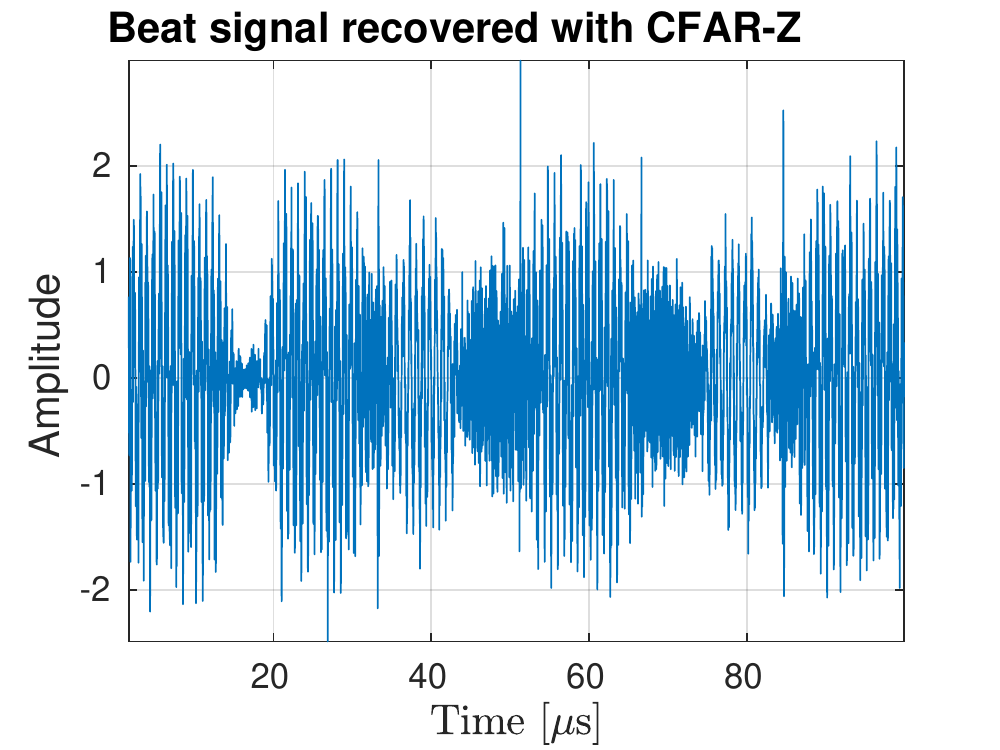}
		\label{fig:pt_RawSig_RealPart_AftSup_15dB_cfarZ}
	}
	\subfloat[]{
		\includegraphics[width=0.325\textwidth]{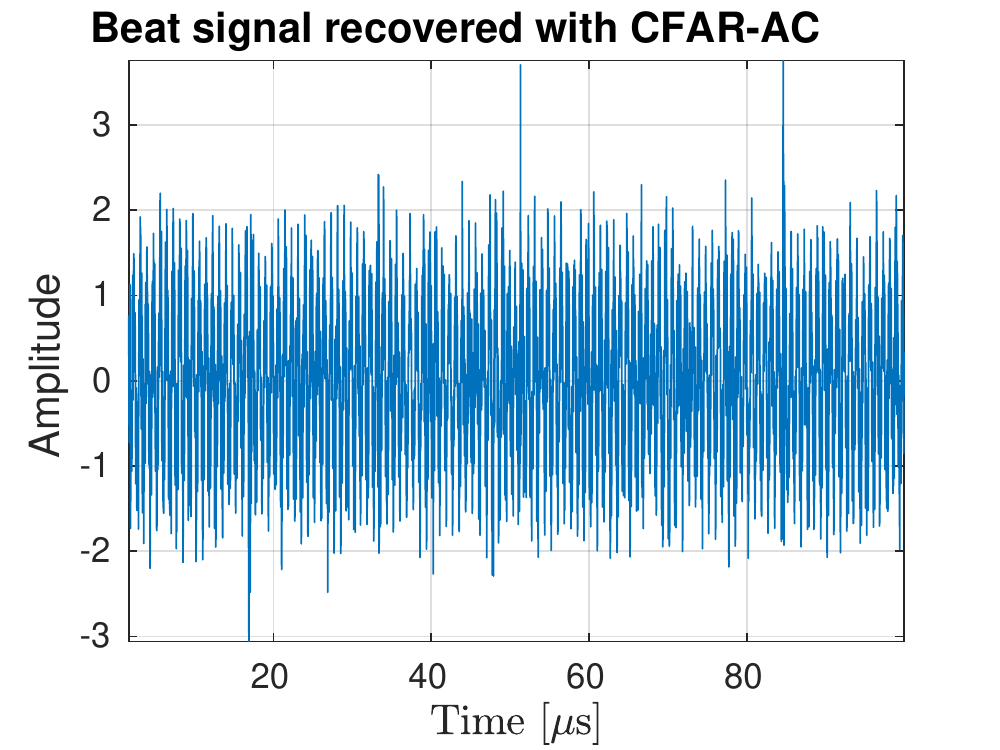}
		\label{fig:pt_RawSig_RealPart_AftSup_15dB_cfarAC}
	}
	\subfloat[]{
		\includegraphics[width=0.325\textwidth]{pt_RawSig_RealPart_AftSup_5dB_ANC}
		\label{fig:pt_RawSig_RealPart_AftSup_ANC}
	}
	
	\vspace{-1mm}
	\subfloat[]{
		\includegraphics[width=0.325\textwidth]{Images/Simu/pt_RawSig_RealPart_AftSup_5dB_WD.pdf}
		\label{fig:pt_RawSig_RealPart_AftSup_15dB_WD}
	}
	\subfloat[]{
		\includegraphics[width=0.325\textwidth]{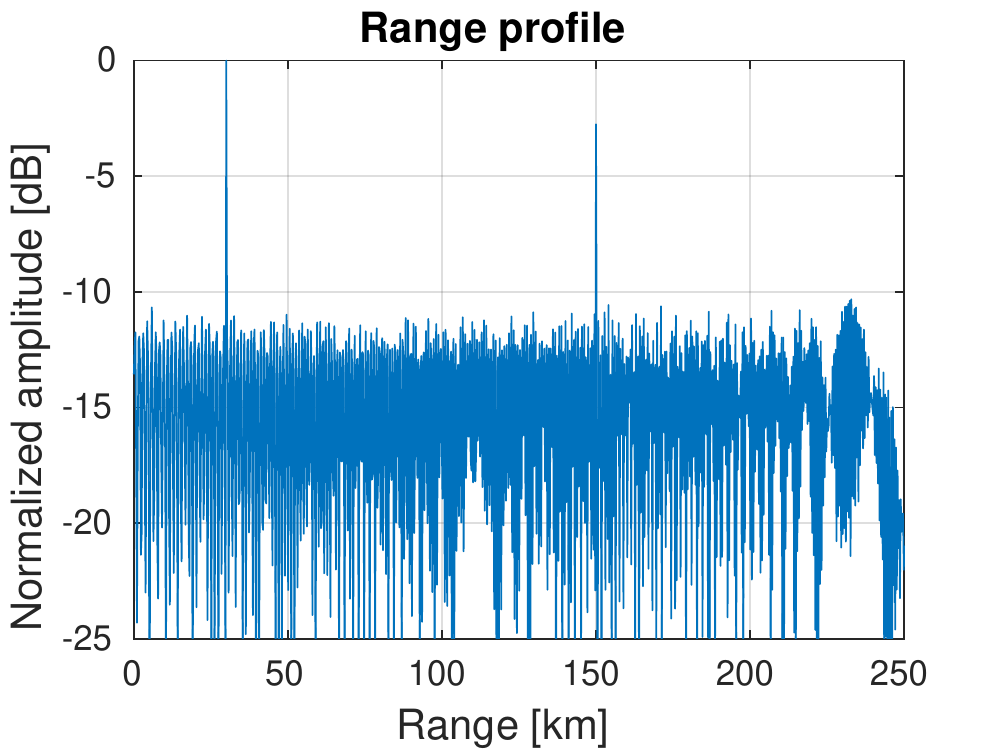}
		\label{fig:pt_RP_15dB}
	}
	\subfloat[]{
		\includegraphics[width=0.325\textwidth]{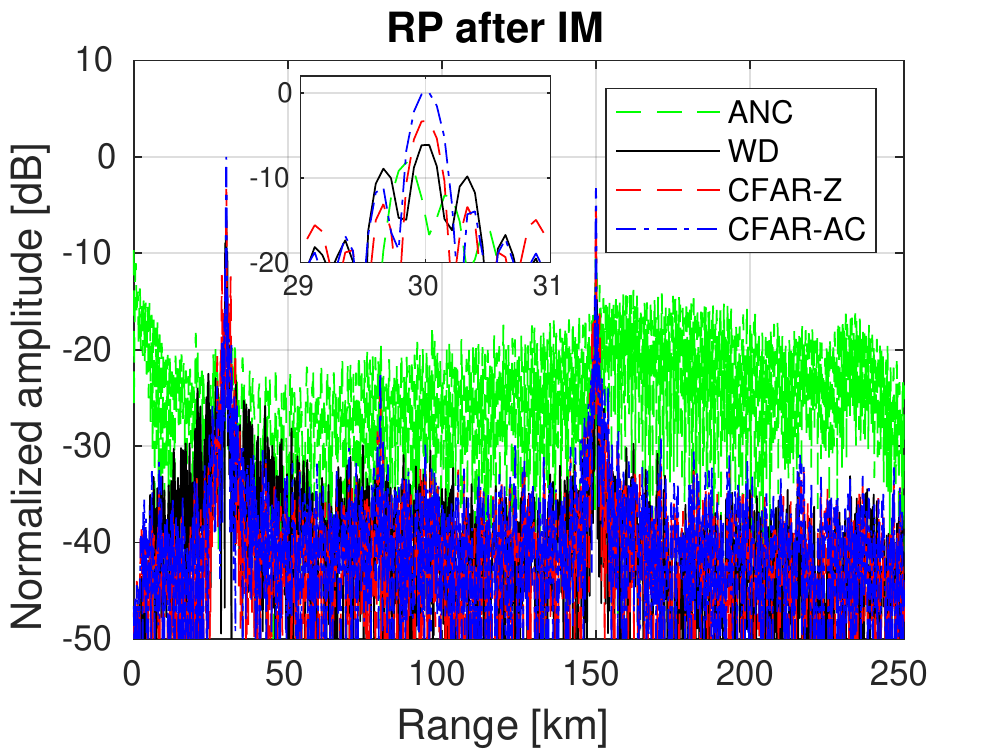}
		\label{fig:pt_RP_AftSup_15dB}
	}

	\caption{Illustration of the CFAR-based interference mitigation and comparisons with the ANC and WD methods. \protect\subref{fig:pt_RawSig_RealPart_15dB} and \protect\subref{fig:pt_TF_15dB} show the real part of the raw signal and its $t$-$f$ spectrum after the STFT. \protect\subref{fig:pt_DetMAP_15dB} displays the map of the detected interferences and \protect\subref{fig:pt_CFAR_mask_15dB} is its dilated version. \protect\subref{fig:pt_TF_15dB_cfarZ} and \protect\subref{fig:pt_TF_15dB_cfarAC} are the $t$-$f$ spectrum after interference mitigation with CFAR-Z and CFAR-AC approaches. \protect\subref{fig:pt_RawSig_RealPart_AftSup_15dB_cfarZ} and \protect\subref{fig:pt_RawSig_RealPart_AftSup_15dB_cfarAC} are the recovered beat signals after taking ISTFT of the spectra in \protect\subref{fig:pt_TF_15dB_cfarZ} and \protect\subref{fig:pt_TF_15dB_cfarAC}. \protect\subref{fig:pt_RawSig_RealPart_AftSup_ANC} and \protect\subref{fig:pt_RawSig_RealPart_AftSup_15dB_WD} show the recovered beat signal after interference mitigation by the ANC and WD methods. \protect\subref{fig:pt_RP_15dB} and \protect\subref{fig:pt_RP_AftSup_15dB} present the range profiles of targets constructed by using the acquired raw signal and the recovered signals after IM, respectively.  }
	\label{fig:Simu_pointTarget_InterfMitig}
\end{figure*}

\begin{figure*}[]
	\centering
	\subfloat[]{
		\includegraphics[width=0.33\textwidth]{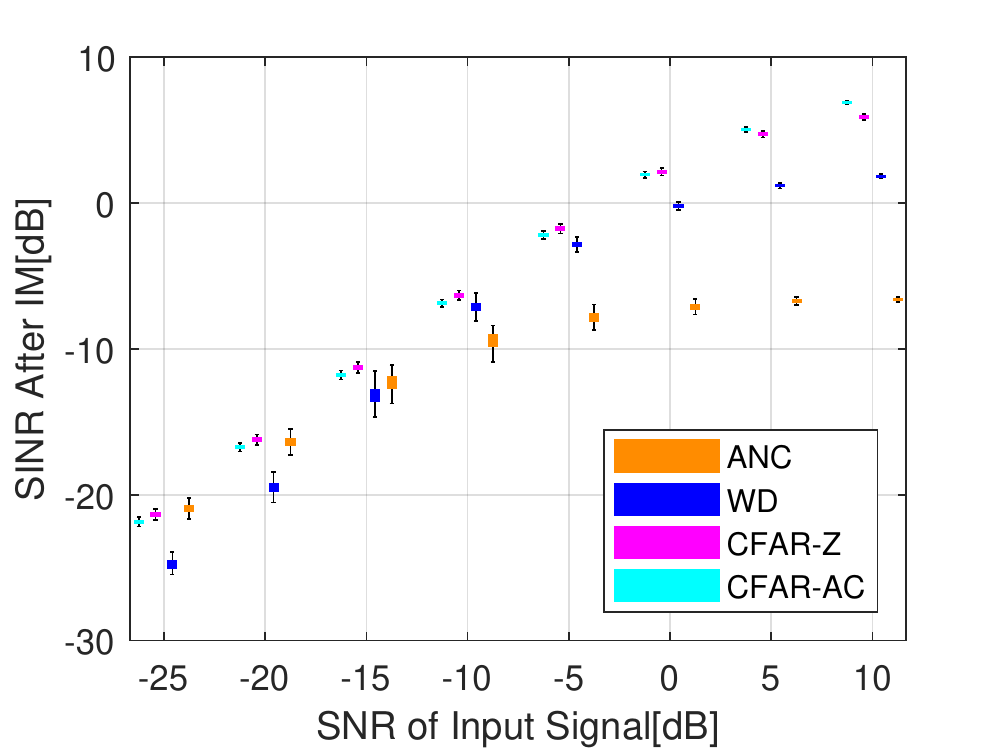}
		\label{fig:simu_performance_SINR}
	}
	\subfloat[]{
		\includegraphics[width=0.33\textwidth]{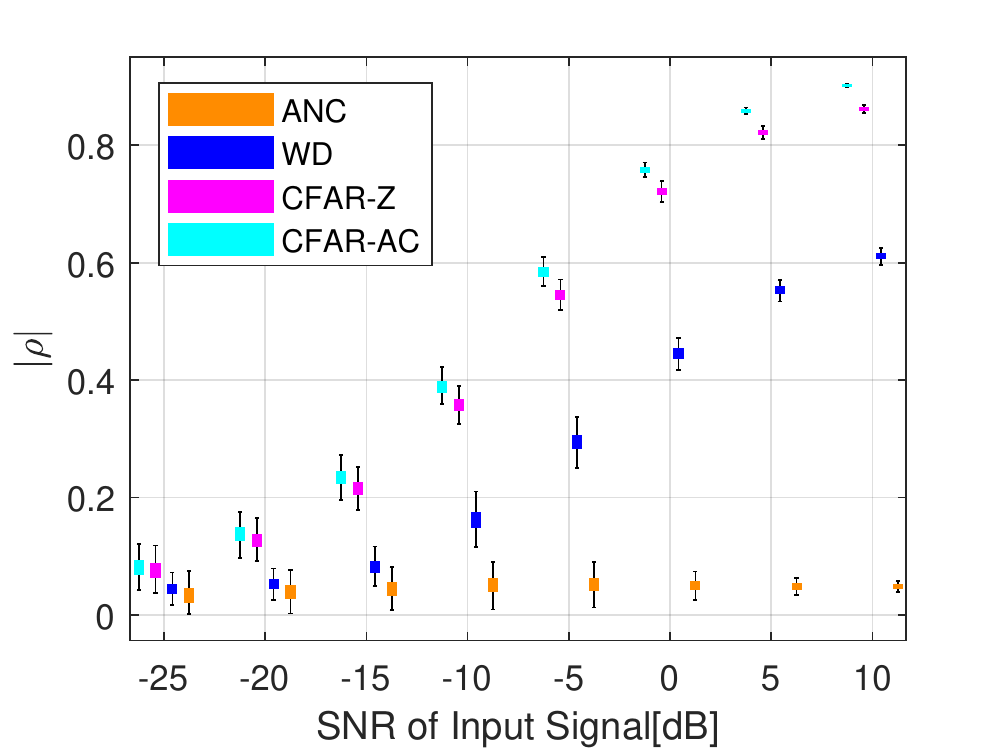}
		\label{fig:simu_performance_rho_abs}
	}
	\subfloat[]{
		\includegraphics[width=0.33\textwidth]{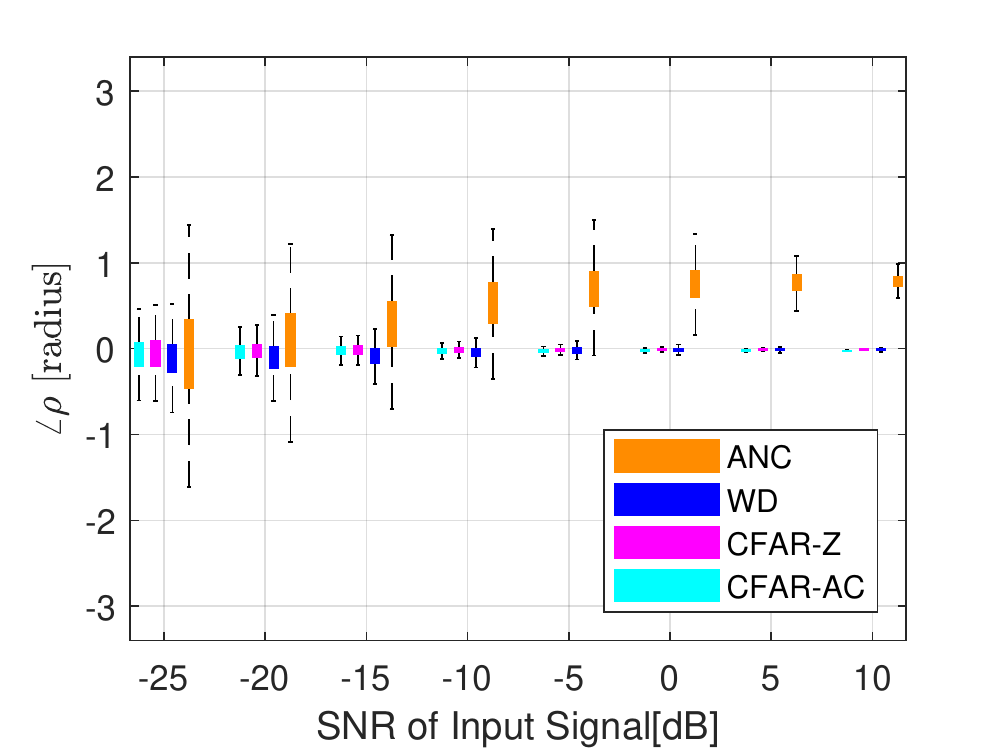}
		\label{fig:simu_performance_rho_phase}
	}
	\caption{Quantitative comparison of the interference mitigation performance of the ANC, WD, CFAR-Z and CFAR-AC methods at the different SNRs of the input signals. \protect\subref{fig:simu_performance_SINR},  \protect\subref{fig:simu_performance_rho_abs} and \protect\subref{fig:simu_performance_rho_phase} show the variations of SINRs, the magnitudes and phase angles of correlation coefficients of the recovered beat signals after interference mitigation, respectively.}
	\label{fig:simu_performance_metric}
\end{figure*}

Numerical simulations are presented to demonstrate the interference mitigation performance of the proposed approach. Meanwhile, the results are compared with two the state-of-the-art efficient approaches, i.e., Wavelet Denoising (WD) approach \cite{Lee2019TITS} and the Adaptive Noise Canceller (ANC) approach \cite{Jin2019TVT}.

\subsection{Performance Metrics}
To facilitate the comparison among different IM approaches and quantitatively evaluate the accuracy of the beat signals recovered by each approach, we use as the metrics the Signal to Interference plus Noise Ratio (SINR) and correlation coefficient ($\rho$) \cite{Wang2020MP} of the beat signal obtained after IM processing relative to the clean reference signal. The SINR is defined in the same way as the relative signal to noise ratio (RSNR) in \cite{Wang2020MP}, which is inversely proportional to the error vector magnitude in \cite{Toth2019RadarConf}. For conciseness, the definition formulas of these metrics are omitted here.      

\subsection{Point Target Simulation} \label{subsec:point_target_simu}

Some typical automotive radar parameters were used for numerical simulations, as listed in Table~\ref{tab:simu_parameters}. Three point targets were placed in the scene of illumination at the distance of $30\,\mathrm{m}$, $80\,\mathrm{m}$ and $150\,\mathrm{m}$, respectively. The amplitudes of the scattered signals from the three targets are set as $1$, $0.1$, and $0.7$ to emulate variations of scattering coefficients of different targets. The victim radar transmitted up-sweep FMCW signals and suffered from some strong FMCW interferences, and complex while Gaussian noise was also added to account for thermal noise and measurement errors of the radar system. The acquired signal at the output of the low-pass filter is shown in Fig.~\ref{fig:Simu_pointTarget_InterfMitig}\subref{fig:pt_RawSig_RealPart_15dB}. Its signal to noise ratio (SNR) and signal to interference plus noise ratio (SINR) are $5\,\mathrm{dB}$ and $-18.71\,\mathrm{dB}$. Due to the strong interferences, the weak target at the distance of $80\,\mathrm{m}$ is completely overwhelmed by the increased noise floor of the range profile formed by taking the FFT of the acquired signal (see Fig.~\ref{fig:Simu_pointTarget_InterfMitig}\subref{fig:pt_RP_15dB}).

Using the proposed approaches to mitigate the interferences, the acquired time-domain signal is first transformed into the $t$-$f$ domain by using the STFT. For discrete implementation, the length of the window of STFT is 256 sampling points and the overlap between adjacent window positions is 252 points. The obtained $t$-$f$ spectrogram is shown in Fig.~\ref{fig:Simu_pointTarget_InterfMitig}\subref{fig:pt_TF_15dB}, where the strong spectrum along the oblique lines are the interferences while the three weak horizontal lines represent the useful beat signals. 

Then, utilizing the CA-CFAR detector along each frequency bin, the non-horizontal patterns of interferences are detected (see Fig.~\ref{fig:Simu_pointTarget_InterfMitig}\subref{fig:pt_DetMAP_15dB}). As the threshold of CA-CFAR detector is computed based on the average of training cells and varies for each Cell Under Test (CUT), it causes the missed detection of the cells at the edges of the oblique lines of the interferences (i.e., the detected lines are thinner than that of the interferences), which leads to only partial mitigation of interferences in the following operations. To overcome this problem, the detection map of the CFAR detector was dilated by using the octagonal structuring element, as shown in Fig.~\ref{fig:Simu_pointTarget_InterfMitig}\subref{fig:pt_CFAR_mask_15dB}. It is clear that the dilated detection map is much thicker compared to that in Fig.~\ref{fig:Simu_pointTarget_InterfMitig}\subref{fig:pt_DetMAP_15dB}. 

Next, the dilated detection map was employed as a mask to zero out the interference-contaminated samples by the CFAR-Z approach or to correct their amplitudes by using the CFAR-AC method. The resultant $t$-$f$ spectra after interference mitigation are shown in Fig.~\ref{fig:Simu_pointTarget_InterfMitig}\subref{fig:pt_TF_15dB_cfarZ} and \subref{fig:pt_TF_15dB_cfarAC}. Finally, applying the ISTFT to the obtained $t$-$f$ spectra, the corresponding beat signals are recovered, shown in Fig.~\ref{fig:Simu_pointTarget_InterfMitig}\subref{fig:pt_RawSig_RealPart_AftSup_15dB_cfarZ} and \subref{fig:pt_RawSig_RealPart_AftSup_15dB_cfarAC}. For comparison, the IM of the signal was also performed using the ANC \cite{Jin2019TVT} and the WD methods \cite{Lee2019TITS}. For the ANC method, the length of the adaptive filter was set 80. Meanwhile, for the WD approach, the level of the wavelet decomposition was four which was optimally selected and the Stein's unbiased risk estimate was used to determine the threshold value. The beat signals recovered by the ANC and the WD methods are presented in Fig.~\ref{fig:Simu_pointTarget_InterfMitig}\subref{fig:pt_RawSig_RealPart_AftSup_ANC} and \subref{fig:pt_RawSig_RealPart_AftSup_15dB_WD}. From Fig.~\ref{fig:Simu_pointTarget_InterfMitig}\subref{fig:pt_RawSig_RealPart_AftSup_15dB_cfarZ}, one can see that the CFAR-Z approach suppresses not only the interferences but also the targets' beat signals at the time instances related to the intersection points of the $t$-$f$ spectra of the interferences and useful signals. Similarly, the wavelet denoising method causes even more loss of useful signals, especially in the period between $35\,\mathrm{\mu s}$ to $85\,\mathrm{\mu s}$ in Fig.~\ref{fig:Simu_pointTarget_InterfMitig}\subref{fig:pt_RawSig_RealPart_AftSup_15dB_WD}. By contrast, the CFAR-AC recovers the beat signals of targets with negligible power loss (Fig.~\ref{fig:Simu_pointTarget_InterfMitig}\subref{fig:pt_RawSig_RealPart_AftSup_15dB_cfarAC}). From Fig.~\ref{fig:Simu_pointTarget_InterfMitig}\subref{fig:pt_RawSig_RealPart_AftSup_ANC}, the ANC method only suppresses part of the interferences between $35\,\mathrm{\mu s}$ and $85\,\mathrm{\mu s}$ and some chirp-like pulses of the interferences are still observed. This could be caused by the fact that the assumption of the complex conjugate symmetry of the interference spectrum around zeros used by the ANC method is not valid to the synthetic data. To quantitatively compare the accuracy of the recovered beat signals relative to the clean reference, the SINRs of the signals obtained with the ANC, WD, CFAR-Z and CFAR-AC methods are $-6.96\,\mathrm{dB}$, $1.27\,\mathrm{dB}$, $4.03\,\mathrm{dB}$ and $6.47\,\mathrm{dB}$, respectively. And the corresponding correlation coefficients are $0.0732e^{-j0.5049}$, $0.5527e^{j0.0167}$, $0.7837e^{-j0.0007}$, and $0.8965e^{0.0297}$.  

Taking the FFT of the recovered beat signals, the targets' range profiles in Fig.~\ref{fig:Simu_pointTarget_InterfMitig}\subref{fig:pt_RP_AftSup_15dB} are obtained. All the approaches except the ANC method significantly suppress the interferences and reduce the noise floor of the focused range profile compared to that in Fig.~\ref{fig:Simu_pointTarget_InterfMitig}\subref{fig:pt_RP_15dB}.  The weak target at the distance of $80\,\mathrm{m}$ is clearly visible. However, compared to that of the WD method, the range profiles obtained with the CFAR-Z and CFAR-AC have lower noise floor and thus achieve better interference mitigation performance. Moreover, in contrast to CFAR-AC approach,  both the WD method and CFAR-Z approach suppress some targets' signals after mitigating the interferences, which not only decreases the signal power but also causes increased sidelobes in the focused range profile (see the inset in Fig.~\ref{fig:Simu_pointTarget_InterfMitig}\subref{fig:pt_RP_AftSup_15dB}). But as mentioned above, the range profile obtained with the CFAR-Z approach still has smaller power loss and lower sidelobes than that acquired with the WD method. Therefore, in terms of noise floor, power loss and sidelobe levels of the resultant range profile, the CFAR-AC achieves the best interference mitigation performance among the three approaches.           

\subsection{Effect of SNR on Interference Mitigation}
The noise included in the acquired signal impacts the detection of interferences, thus affecting the interference mitigation. In this section, we used the same targets' signals and the interferences as in section~\ref{subsec:point_target_simu} but changed the added noise levels to investigate the effect of SNR on the IM performance of the two proposed approaches and their competing counterparts, i.e., WD and ANC methods. 

The noise levels with the SNR ranging from $-25\,\mathrm{dB}$ to $10\,\mathrm{dB}$ were considered. At each noise level, 500 times Monte Carlo runs were implemented and the statistics of the performance metrics achieved by the four IM methods are presented as the box plot in Fig.~\ref{fig:simu_performance_metric}. The bottom and top of each box indicate the $25^\text{th}$ and $75^\text{th}$ percentiles of the sample, respectively. Meanwhile, the lines extending above and below of each box show the range between the maximum and minimum values of the sample. From Fig.~\ref{fig:simu_performance_metric}\subref{fig:simu_performance_SINR}, one can see that the SINR of the recovered signals after IM increases with the increase of the SNR of input signal. Generally, the proposed CFAR-Z and CFAR-AC achieve better SINR than the WD and ANC approaches except at $\text{SNR}=-25\,\mathrm{dB}$ in which case the interferences are almost overwhelmed by the noise. A large portion of the interferences were not detected by the CFAR-Z and CFAR-AC approaches; as a results, the interferences are not fully suppressed. Meanwhile, the WD method also fail to extra the interferences and leads to degraded SINR after IM. By contrast, the ANC method eliminates half of the frequency spectrum that does not contain targets' signals and uses the complex conjugate symmetry of the interference spectra to suppress them; thus, it results in better SINR after IM. In addition, compared to CFAR-AC, the CFAR-Z obtains slightly higher SINRs when $\text{SNR} < 0\,\mathrm{dB}$ but lower ones when $\text{SNR} > 0\,\mathrm{dB}$. This is because that when $\text{SNR}<0\,\mathrm{dB}$ the values of targets' signals at the interference-contaminated region are more closer to zero than to the amplitude-corrected values in which noise is the dominant component. Therefore, in terms of the SINR obtained after IM, the CFAR-Z approach is a better option than the CFAR-AC when the SNR of the input signal is lower than $0\,\mathrm{dB}$.    

Moreover, Fig.~\ref{fig:simu_performance_metric}\subref{fig:simu_performance_rho_abs} shows that the magnitudes of correlation coefficients of the signals obtained with the CFAR-AC are constantly larger than that acquired by the other three methods. Moreover, with the rise of the SNRs of input signals, the phase angles of the correlation coefficients of the recovered signals by the WD, CFAR-Z and CFAR-AC are all increasingly concentrated around zero. So in terms of both the SINR and correlation coefficient of the recovered signals after IM, the proposed CFAR-Z and CFAR-AC outperform the other two approaches; however, in practice the better choice between them should be determined based on the SNR of the acquired signal.       

\subsection{Computational Time}

{
	\renewcommand{\arraystretch}{1.2}
\begin{table}[!t]
	\centering
	\caption{Computational time of the ANC, WD, CFAR-Z and CFAR-AC approaches for interference mitigation}
	\label{tab:simu_computationTime}
	\begin{threeparttable}
	\begin{tabular}{l|c|c|c|c|c|c}
		\hline
		\multirow{2}{*}{} & \multirow{2}{*}{ANC} & \multirow{2}{*}{WD} & \multicolumn{2}{c|}{CFAR-Z}       & \multicolumn{2}{c}{CFAR-AC}      \\ \cline{4-7} 
		&                      &                     & $N_1^\text{os}$ \tnote{1} &  $N_2^\text{os}$  & $N_1^\text{os}$ & $N_2^\text{os}$ \\ \hline
		Time {[}ms{]}     & 73                   & 9.64                & 214             & 109.3           & 228.5           & 116.4           \\ \hline
	\end{tabular}
	\begin{tablenotes}
		\item \footnotesize{$^1$$N_1^\text{os}=252$, $N_2^\text{os}=248$} are the number of overlapped samples of the sliding window at two adjacent positions for the STFT. 
	\end{tablenotes}
	\end{threeparttable}
\end{table}
}
The computational complexities of the proposed CFAR-Z and CFAR-AC are dominated by the CFAR detection along each frequency bin. In section.~\ref{subsec:point_target_simu}, the synthetic beat signal in one sweep contains 3933 samples and was processed with the ANC, WD, CFAR-Z and CFAR-AC approaches by using MATLAB 2019b on a computer with Intel i5-3470 CPU and $8\,\mathrm{GB}$. The computational time of the four IM approaches are summarized in Table~\ref{tab:simu_computationTime}. One can see that the WD method is the most efficient one compared to the other three approaches. Meanwhile, when the number of overlapped samples of the STFT window decreases from 252 to 248 (i.e., the sliding step of the STFT window increases from 4 to 8), the computational time of both CFAR-Z and CFAR-AC decreases by about $50\%$ as the number of samples in each frequency bin is halved for the CFAR detection. So by properly adjusting the processing parameter, the computational time of the CFAR-Z and CFAR-AC can be significantly reduced. Moreover, as the CFAR detection is carried out independently along each frequency bin in the $t$-$f$ domain, these detection operations along different frequency bins can be implemented by using parallel computing, thus further reducing their computational time and improving the real-time processing capability.

\section{Experimental results}  \label{sec:exp_result}

\begin{figure}[!t]
	\centering
	\subfloat[]{
		\includegraphics[width=0.40\textwidth]{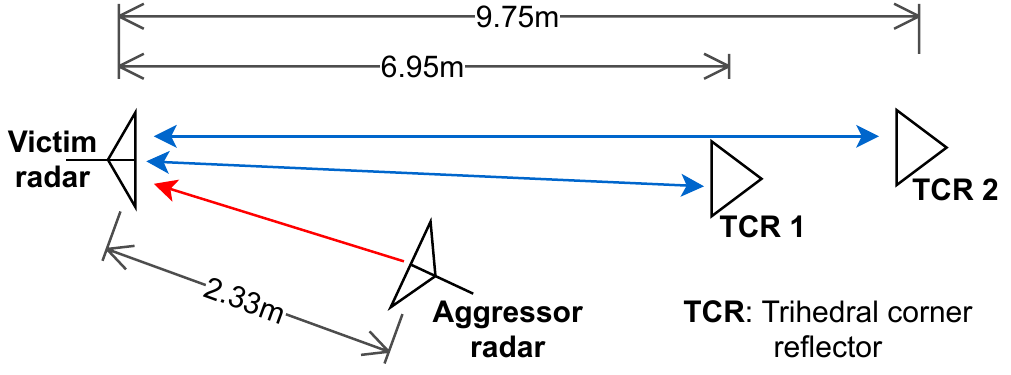}
		\label{fig:exp_setup_geometry}
	}

	\vspace{-2mm}
	\subfloat[]{
		\includegraphics[width=0.35\textwidth]{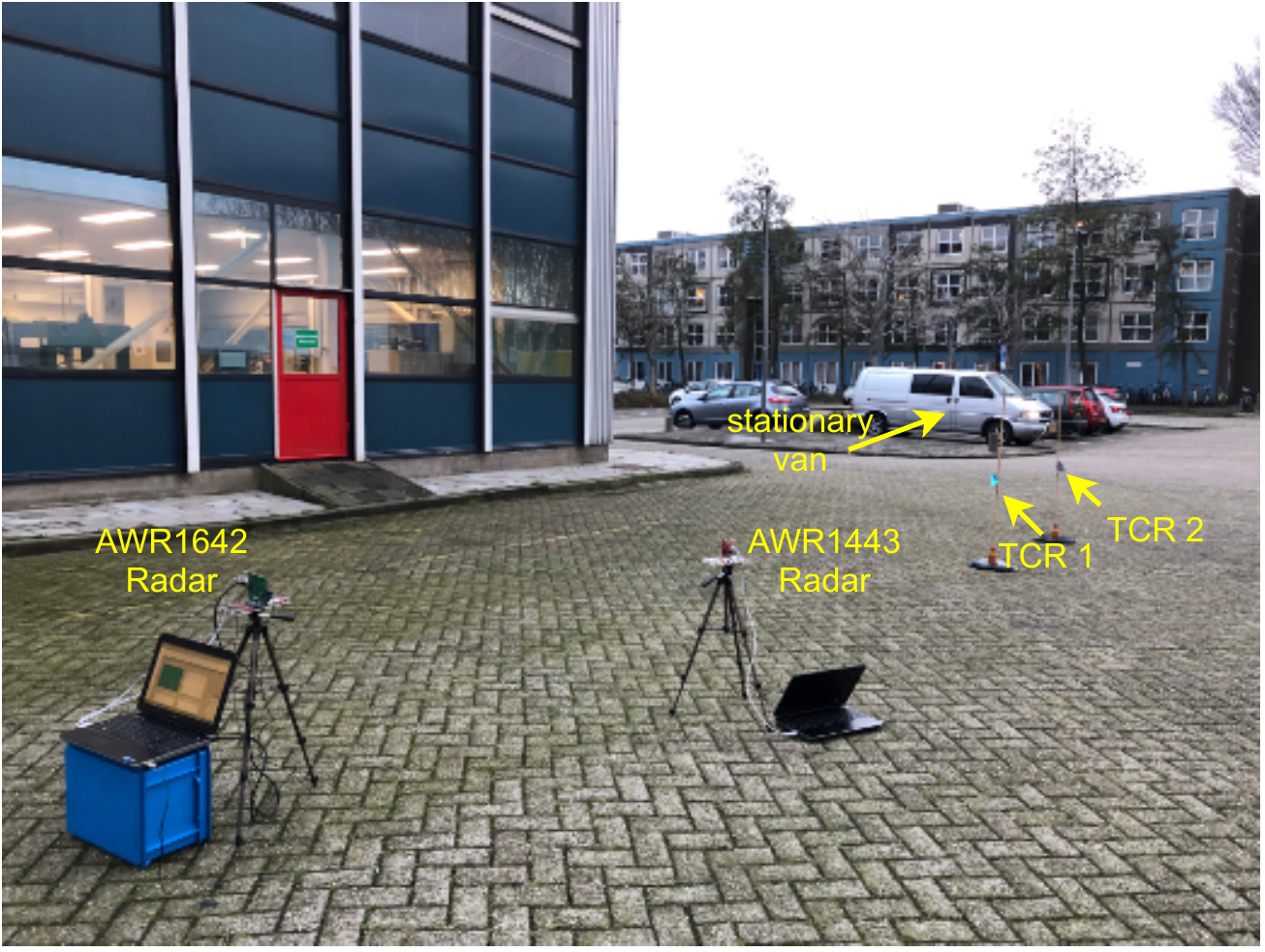}
		\label{fig:exp_setup_pic}
	}
	\caption{Experimental setup for interference mitigation with two TI automotive radar boards. \protect\subref{fig:exp_setup_geometry} shows geometrical configuration and \protect\subref{fig:exp_setup_pic} the experimental setup.}  
	\label{fig:exp_setup}
\end{figure}

{
	\renewcommand{\arraystretch}{1.2}
	\begin{table}[!t]
		\caption{Parameters of experimental radar systems}
		\label{tab:exp_radar_parameter}
		\begin{tabular}{c|c|c|l}
			\hline
			\multicolumn{1}{c|}{\textbf{Parameter}} & \multicolumn{1}{c|}{\textbf{Victim radar}} & \multicolumn{1}{c|}{\textbf{Aggressor radar}} & \textbf{Unit} \\ \hline
			Center frequency                         & 77.69                                       & 77.69                                          & $\mathrm{GHz}$           \\ \hline
			Bandwidth                                & 1380.18                                        &    1380                                           & $\mathrm{MHz}$           \\ \hline
			K                                        & 15.015                                      &     35                                          & $\mathrm{MHz/ \mu s}$ \\ \hline
			T                                        & 91.92                                        &    39.4337                                           & $\mu s$       \\ \hline
			Sampling frequency                        & 6.25                                          &  5                                             & $\mathrm{MHz}$           \\ \hline 
			No. of Samples &   512 &  256 &  -- \\ \hline
		\end{tabular}
	\end{table}
}

\begin{figure*}[!t]
	\centering
	\subfloat[]{
		\includegraphics[width=0.33\textwidth]{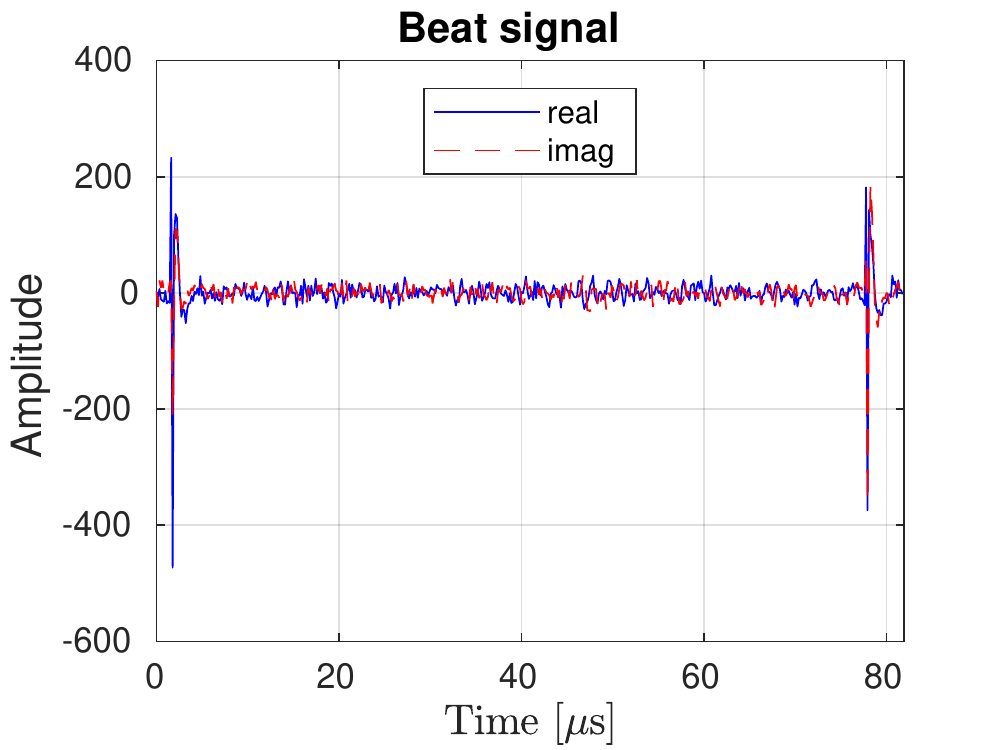}
		\label{fig:exp_RawSig_RealPart}
	}
	\subfloat[]{
		\includegraphics[width=0.33\textwidth]{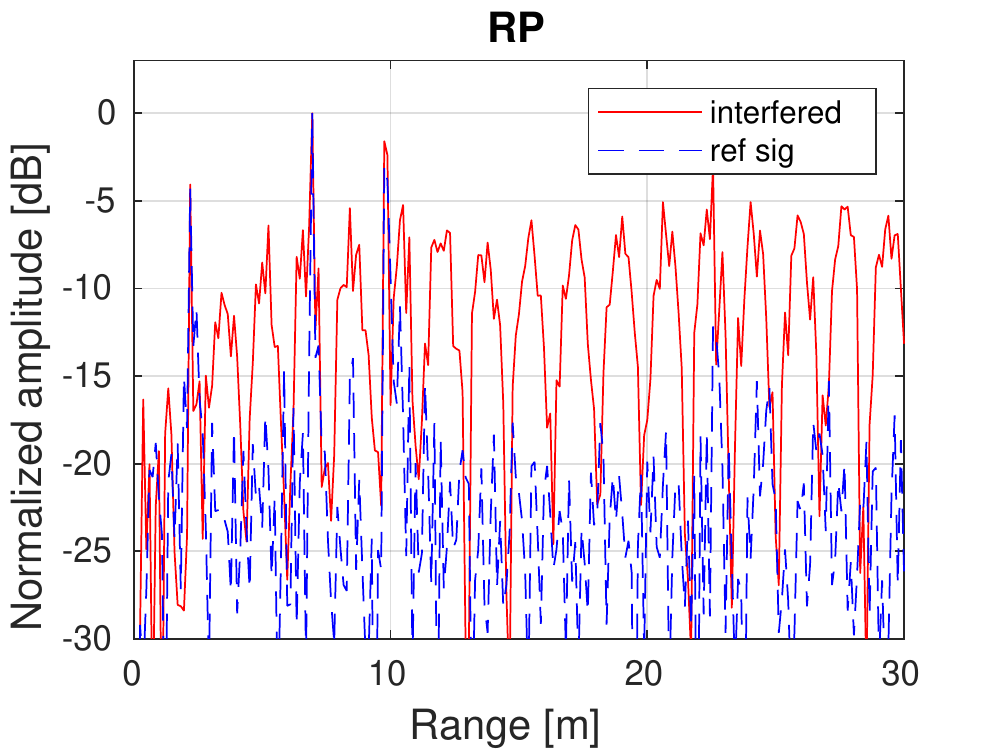}
		\label{fig:exp_RP}
	}	
	\subfloat[]{
		\includegraphics[width=0.33\textwidth]{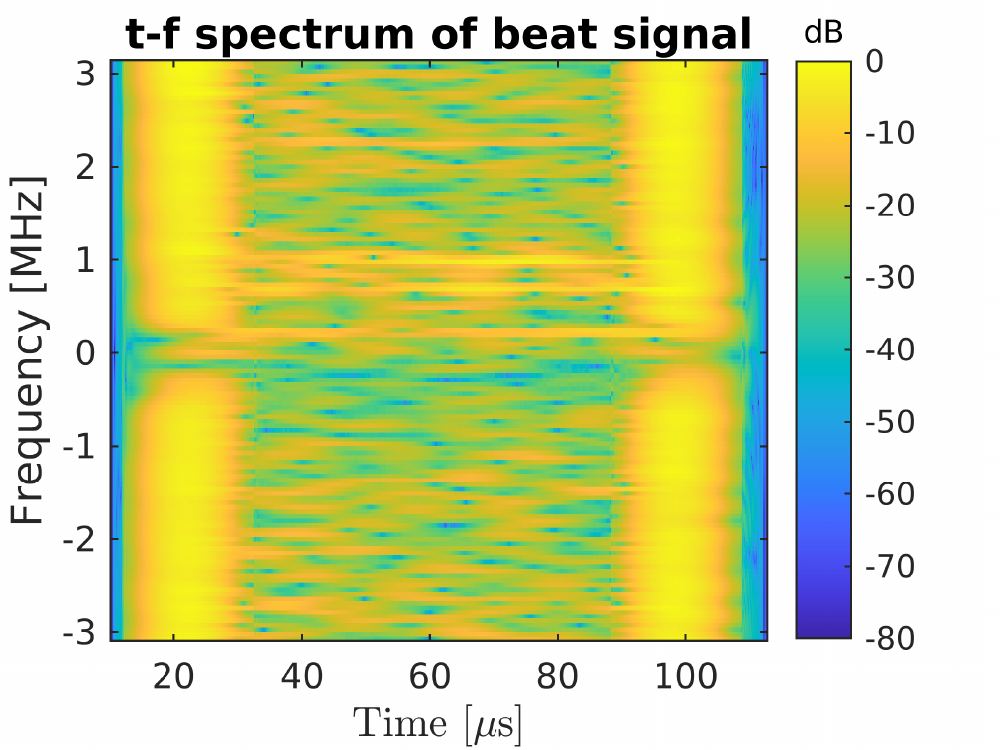}
		\label{fig:exp_TF}
	}

	\subfloat[]{
		\includegraphics[width=0.33\textwidth]{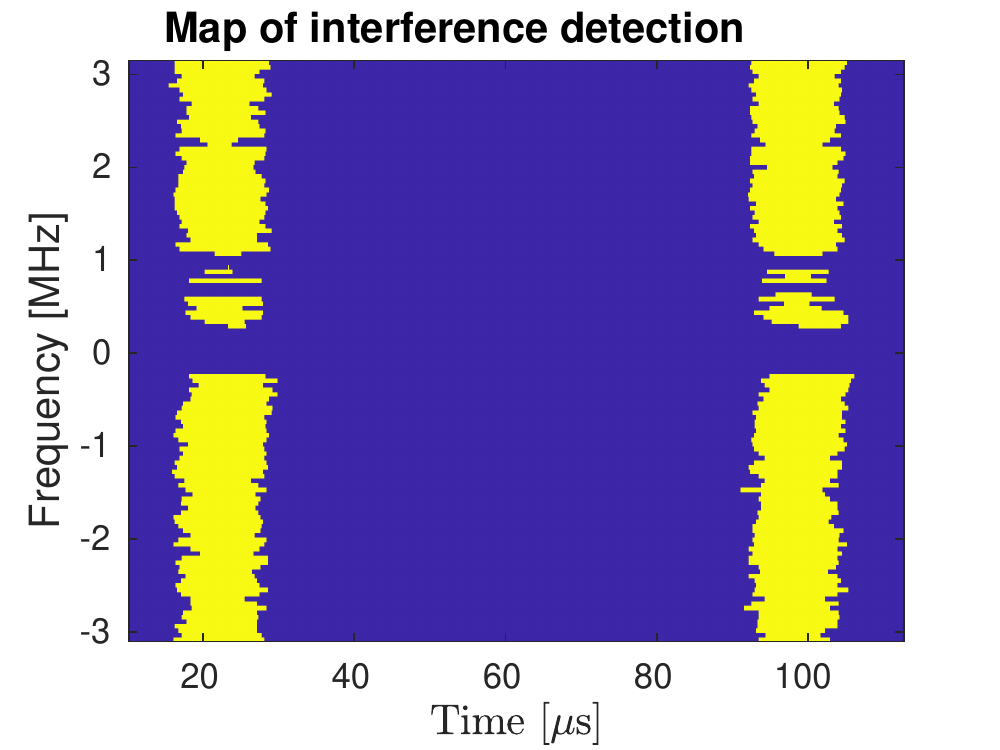}
		\label{fig:exp_DetMap}
	}
	\subfloat[]{
		\includegraphics[width=0.33\textwidth]{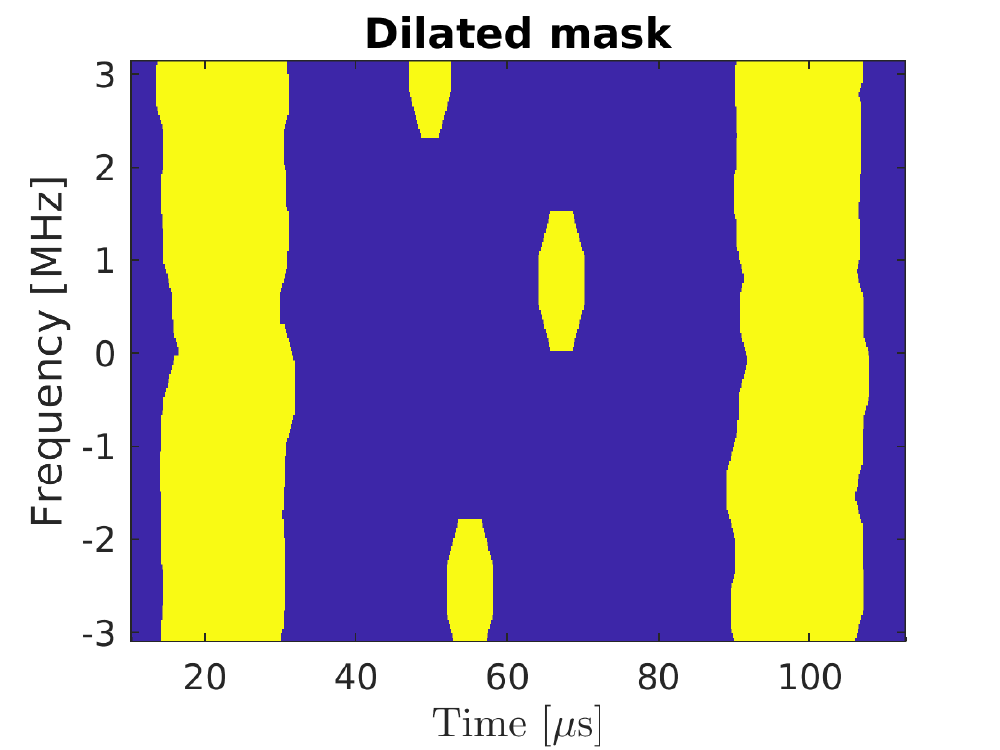}
		\label{fig:exp_CFAR_mask}
	}
	\subfloat[]{
		\includegraphics[width=0.33\textwidth]{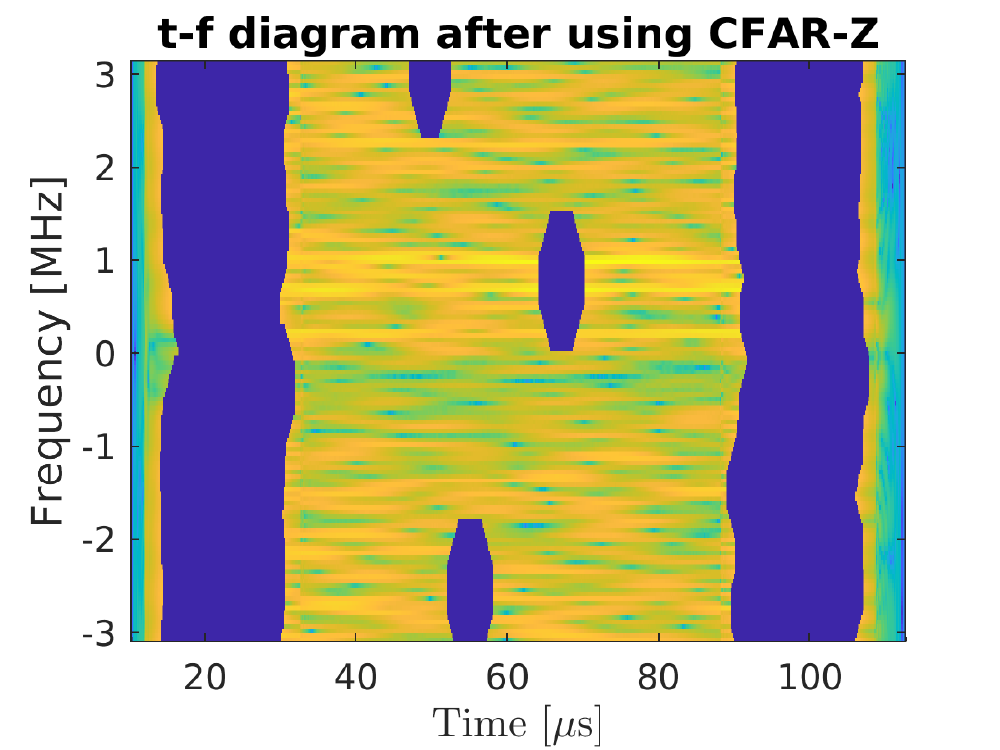}
		\label{fig:exp_TF_cfarz}
	}

	\subfloat[]{
		\includegraphics[width=0.33\textwidth]{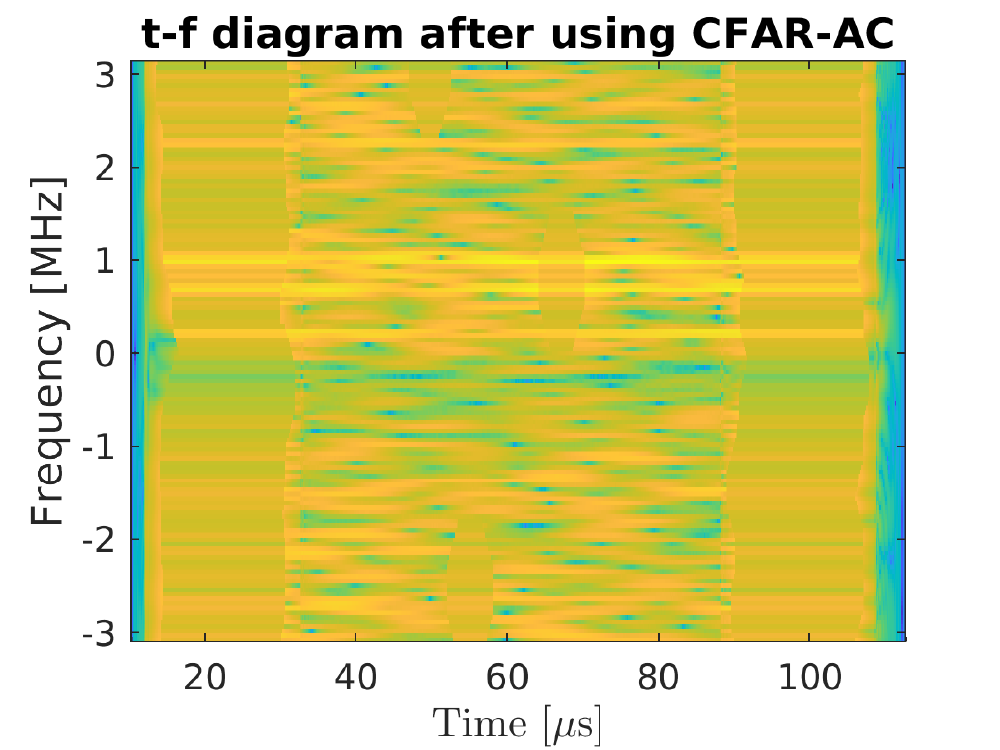}
		\label{fig:exp_TF_cfarAC}
	}
	\subfloat[]{
		\includegraphics[width=0.33\textwidth]{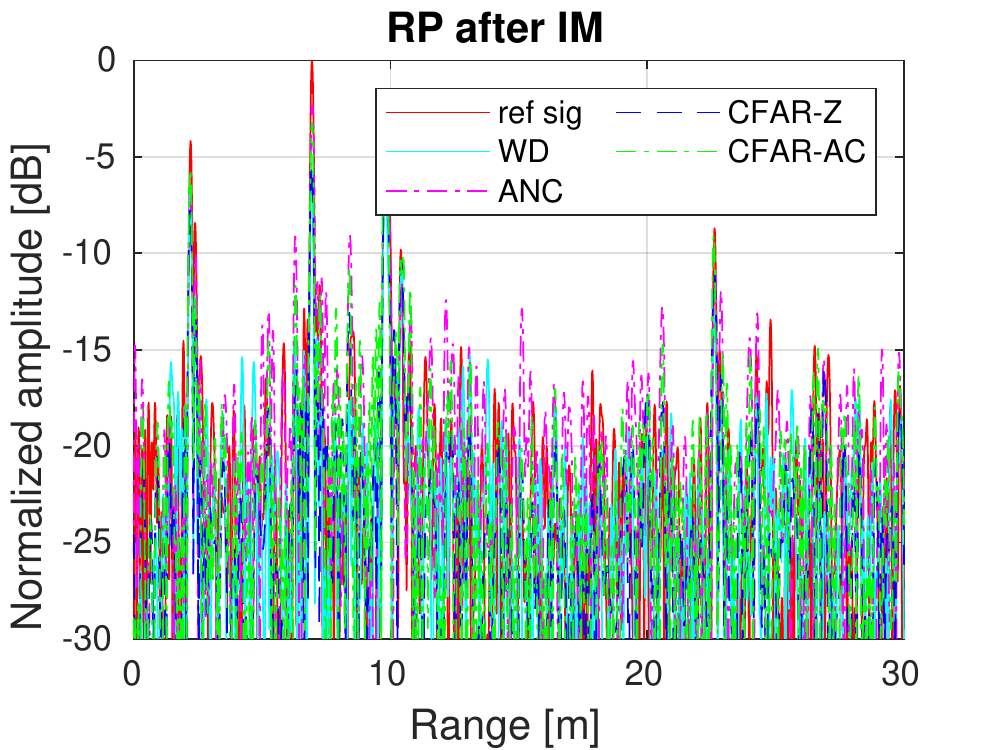}
		\label{fig:exp_RP_AftSup}
	}
	\subfloat[]{
		\includegraphics[width=0.33\textwidth]{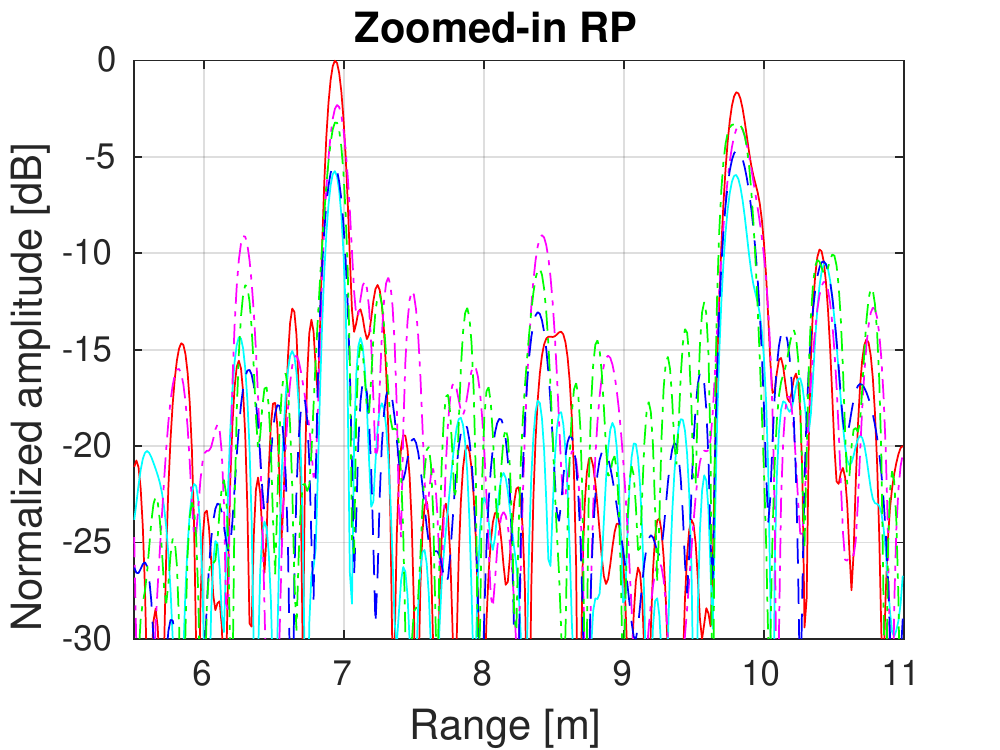}
		\label{fig:exp_RP_AftSup_zoom}
	}
	\caption{Interference mitigation for the experimental radar measured with TI automotive radar. \protect\subref{fig:exp_RawSig_RealPart} shows the acquired beat signal contaminated by the interferences and \protect\subref{fig:exp_TF} its time-frequency spectrum. \protect\subref{fig:exp_RP} presents the range profiles of targets related to the beat signal in \protect\subref{fig:exp_RawSig_RealPart} and an interference-free reference. \protect\subref{fig:exp_DetMap} shows the CFAR detection map of the interferences and \protect\subref{fig:exp_CFAR_mask} its dilation that will be used for interference mitigation. \protect\subref{fig:exp_TF_cfarz} and \protect\subref{fig:exp_TF_cfarAC} give the results of interference mitigation with CFAR-Z and CFAR-AC approaches, respectively. \protect\subref{fig:exp_RP_AftSup} displays the range profiles obtained after interference mitigation and \protect\subref{fig:exp_RP_AftSup_zoom} shows the zoomed-in view of the range profiles at the distance of $5.5\,\mathrm{m}$ to $11\,\mathrm{m}$.}
	\label{fig:exp_IM_TI_radar}
\end{figure*}

\begin{figure*}[]
	\centering
	\subfloat[]{
		\includegraphics[width=0.33\textwidth]{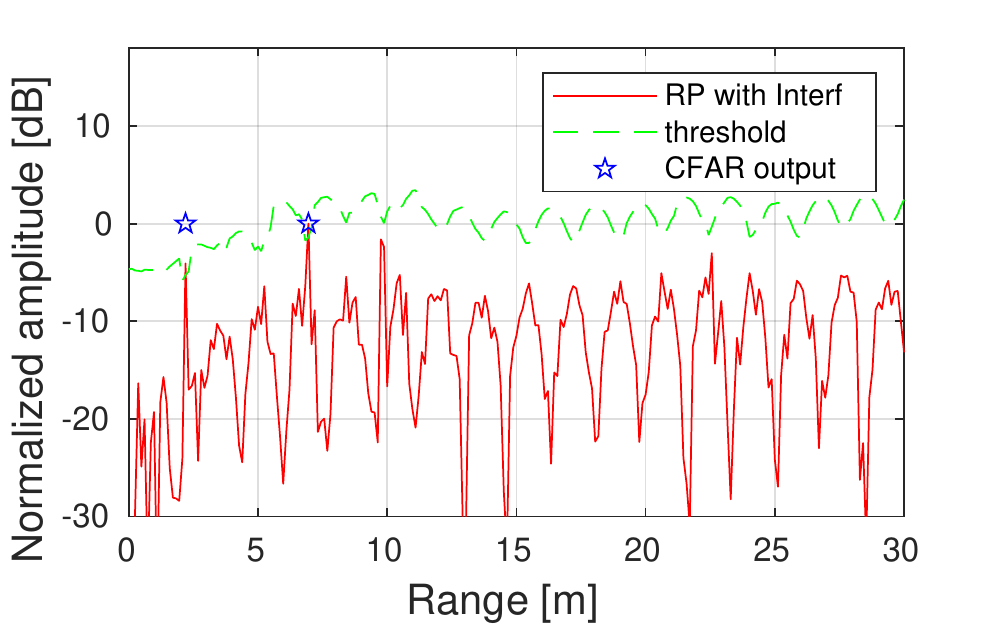}
		\label{fig:exp_RP_detec_BefIM}
	}
	\subfloat[]{
		\includegraphics[width=0.33\textwidth]{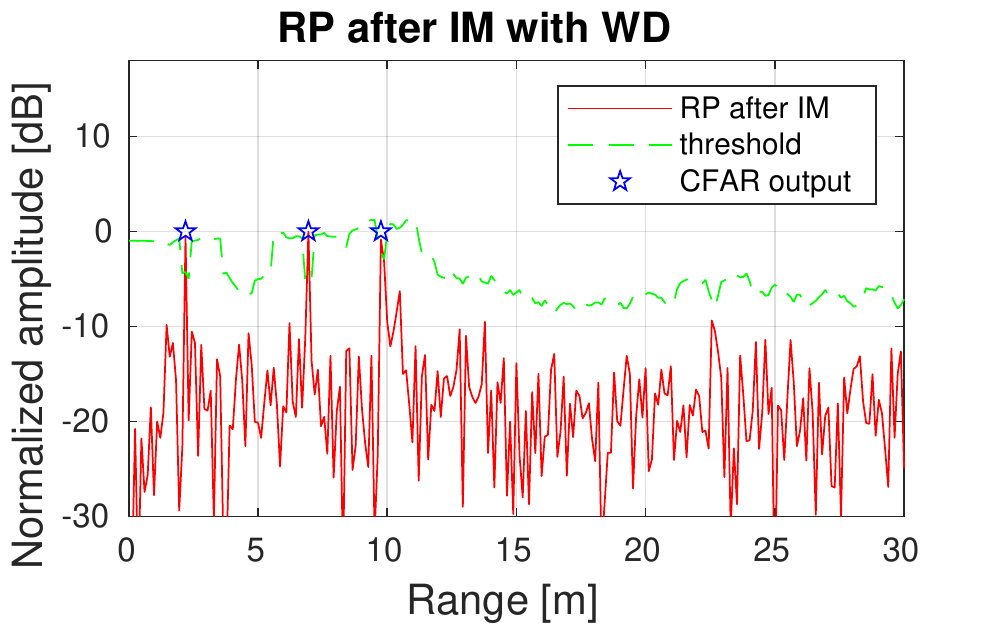}
		\label{fig:exp_RP_detec_AftIM_wd}
	}
	\subfloat[]{
		\includegraphics[width=0.33\textwidth]{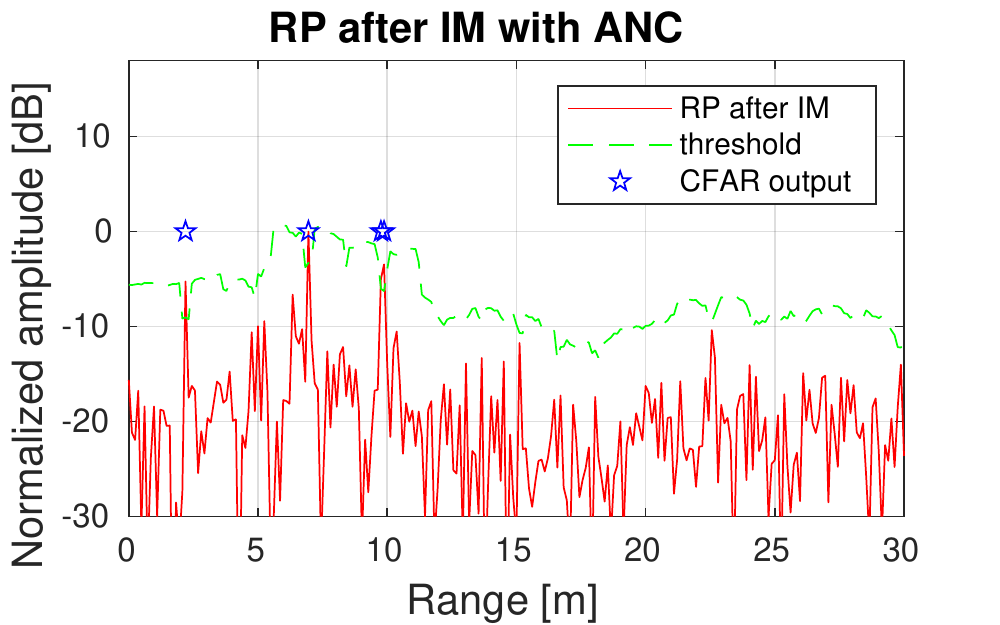}
		\label{fig:exp_RP_detec_AftIM_ANC}
	}

	\subfloat[]{
		\includegraphics[width=0.33\textwidth]{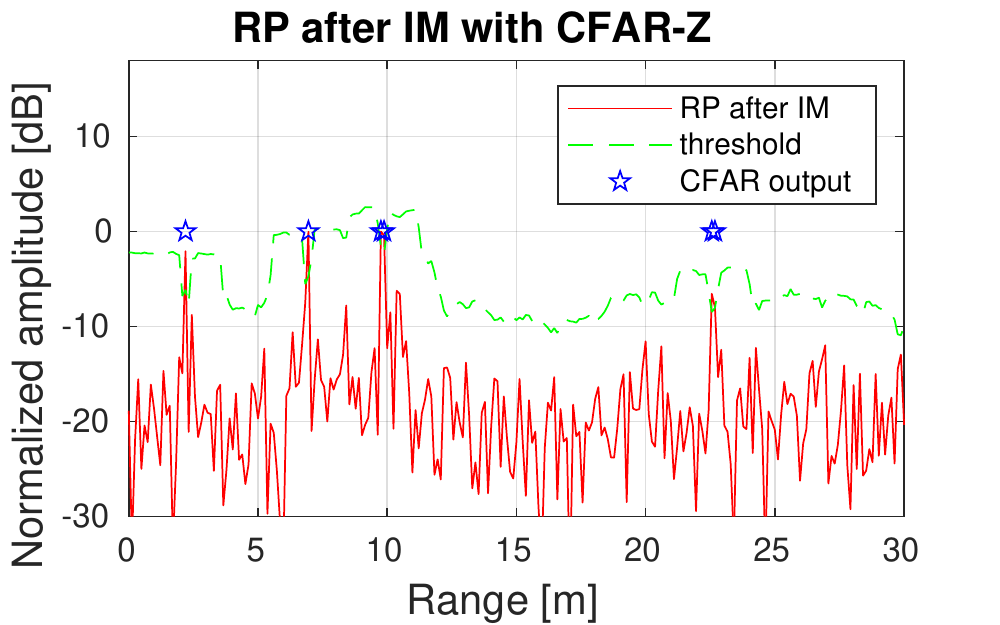}
		\label{fig:exp_RP_detec_AftIM_CFAR_Z}
	}
	\subfloat[]{
		\includegraphics[width=0.33\textwidth]{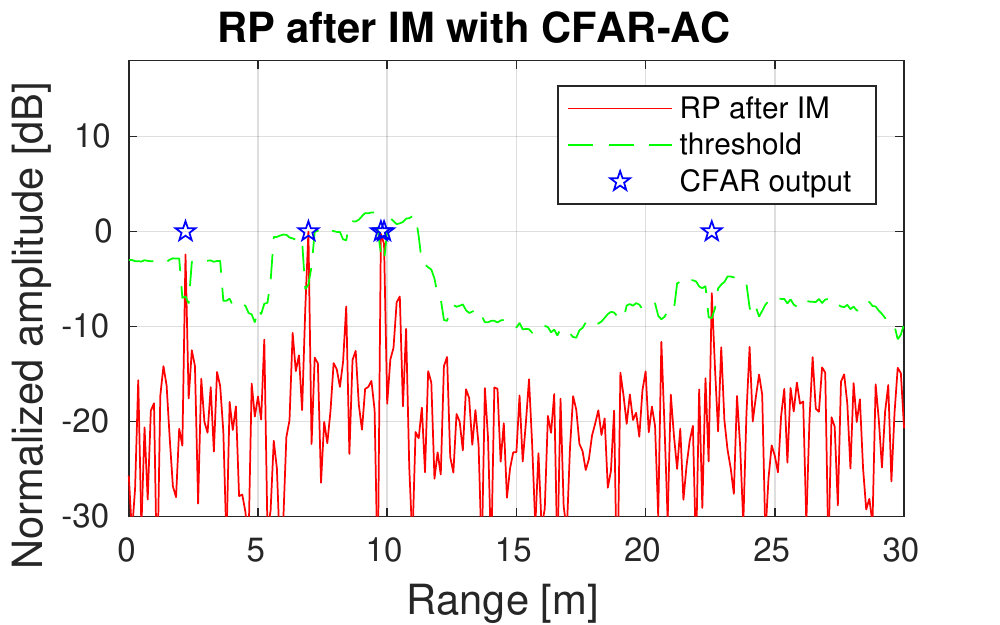}
		\label{fig:exp_RP_detec_AftIM_CFAR_AC}
	}
	
	\caption{The output results of the CFAR detection of the input range profiles before and after interference mitigation. \protect\subref{fig:exp_RP_detec_BefIM} shows the detected results of the range profiles obtained with interference-contaminated signal. \protect\subref{fig:exp_RP_detec_AftIM_wd}, \protect\subref{fig:exp_RP_detec_AftIM_ANC}, \protect\subref{fig:exp_RP_detec_AftIM_CFAR_Z} and \protect\subref{fig:exp_RP_detec_AftIM_CFAR_AC} are the corresponding detection results after taking interference mitigation with WD, ANC, CFAR-Z and CFAR-AC approaches, respectively.}
	\label{fig:exp_RP_detection}
\end{figure*}

In this section, the experimental results are presented to demonstrate the performance of the proposed approach.

One Texas Instruments (TI) AWR1642BOOST radar board is used as the victim radar while another TI AWR1443BOOST radar board is utilized as the aggressor radar. Two Trihedral Corner Reflectors (TCRs) are used as targets. The geometrical configuration and picture of experimental setup are shown in Fig.~\ref{fig:exp_setup}\subref{fig:exp_setup_geometry} and \subref{fig:exp_setup_pic}. The system parameters used for the victim and aggressor radars are listed in Table~\ref{tab:exp_radar_parameter}. The AWR1642 radar board is connected with a TI DCA1000EVM data capture card to collect raw ADC data which is then sent to a host laptop for data storage.

Fig.~\ref{fig:exp_IM_TI_radar}\subref{fig:exp_RawSig_RealPart} shows the acquired signal in one of the FMCW sweeps.  Two large pulses are observed at the the beginning and end of the acquired data, which are caused by the strong interferences from the aggressor radar. After range compression, the interference-contaminated signal leads to a range profile with significantly increased noise floor (see Fig.~\ref{fig:exp_IM_TI_radar}\subref{fig:exp_RP}). For comparison, the range profile of targets obtained with a clean reference signal is also presented, where the first three peaks indicate the locations of the aggressor radar and two TCRs at the distance of $2.33\,\mathrm{m}$, $6.95\,\mathrm{m}$ and $9.75\,\mathrm{m}$, respectively. One can observe that in the range profile obtained with the interference-contaminated signal the two TCRs are almost overwhelmed by the increased noise floor. This made the TCR at the further distance not detectable when a CA-CFAR detector was employed for target detection (see Fig.~\ref{fig:exp_RP_detection}\subref{fig:exp_RP_detec_BefIM}).  The CA-CFAR detector was set with one guard cell and 10 training cells on each side of the CUT and the probability of false alarm of $1\times 10^{-4}$.

To overcome the missed detection of the target caused by the strong interferences, the proposed approaches are applied to the acquired signal for interference mitigation. Firstly, the $t$-$f$ spectrum of the acquired signal is computed through the STFT implemented by using a sliding Hamming window of length 128 with sliding step of one for signal segmentation and then taking the FFT of each signal segments. The obtained spectrum is shown in Fig.~\ref{fig:exp_IM_TI_radar}\subref{fig:exp_TF}. One can see that the interferences exhibit as the two thick vertical lines in the $t$-$f$ domain. Then, a 1-D CFAR detector is applied along each frequency bin to detect the interference-contaminated signal spectrum, and the detection map is presented in   
Fig.~\ref{fig:exp_IM_TI_radar}\subref{fig:exp_DetMap}. Compared to Fig.~\ref{fig:exp_IM_TI_radar}\subref{fig:exp_TF}, one can see that in Fig.~\ref{fig:exp_IM_TI_radar}\subref{fig:exp_DetMap} some interference-contaminated spectral samples are not detected, especially the ones at the edges of the two thick spectral lines. To tackle the missed detection of the interferences, the detection map is dilated with an octagonal structuring element and the result is displayed in Fig.~\ref{fig:exp_IM_TI_radar}\subref{fig:exp_CFAR_mask}. Note that three small patches appear between $40\,\mathrm{\mu s}$ and $80\,\mathrm{\mu s}$, which reveals that some isolated spectral samples are falsely detected as the interference in Fig.~\ref{fig:exp_IM_TI_radar}\subref{fig:exp_DetMap}. Next, using the dilated detection map as a mask, the zeroing and amplitude correction can be conducted to substantially mitigate the inferences, and the results of CFAR-Z and CFAR-AC approaches are given in Fig.~\ref{fig:exp_IM_TI_radar}\subref{fig:exp_TF_cfarz} and \subref{fig:exp_TF_cfarAC}. Finally, the  $t$-$f$ spectrum obtained after IM is inverted through the ISTFT to reconstruct the time-domain beat signal. 

To demonstrate the IM performance of the proposed approaches, the targets' range profiles resulting from their recovered beat signals are presented in Fig.~\ref{fig:exp_IM_TI_radar}\subref{fig:exp_RP_AftSup}. For comparison, the range profiles obtained with the reference signal and the beat signals recovered by the two IM methods, i.e. WD and ANC, are also presented, which are normalized by the maximum value of all the range profiles. From Fig.~\ref{fig:exp_IM_TI_radar}\subref{fig:exp_RP_AftSup}, the overall range profiles obtained with the WD, CFAR-Z and CFAR-AC have very good agreement with the reference one except that the one acquired by the ANC method has higher sidelobes. However, based on the zoomed-in view of the range profiles around the two TRCs (Fig.~\ref{fig:exp_IM_TI_radar}\subref{fig:exp_RP_AftSup_zoom}), one can see that among the four IM methods, the CFAR-AC and ANC methods get the maximum peak values at the distances of two TCRs, whose values are also closest to the reference ones. However, the WD method results in lower peak amplitudes than the reference one and the other three methods as the wavelet-based denoising method not only eliminates the strong interferences but also suppresses part of the useful signal power. Meanwhile, as expected, the CFAR-Z leads to smaller peak values of the range profile at the positions of two TCRs compared to the CFAR-AC. So in terms of power conservation of useful signals, the CFAR-AC and ANC methods achieve the best performance in this case. However, the ANC method assumes strict complex conjugate symmetry of the interference spectrum in the positive and negative frequency bands. Otherwise, its performance degrades significantly as shown in the simulation. In addition, we want to mention that to avoid the weighting effect of the sliding window of the STFT on the reconstructed signal samples in the beginning and the end, 128 zeros were padded at both sides of the acquired signal before computing the STFT and then the extra zeros were removed after inverting the $t$-$f$ spectrum through the ISTFT. Due to this operation before the STFT, it leads to the visually ``increased'' time duration of the $t$-$f$ domain plots (i.e., Fig.~\ref{fig:exp_IM_TI_radar}\subref{fig:exp_TF}-\subref{fig:exp_TF_cfarAC}) compared to that of the acquired signal (Fig.~\ref{fig:exp_IM_TI_radar}\subref{fig:exp_RawSig_RealPart}).         

To further evaluate the quality of the beat signals recovered by the four IM approaches, the target detection performance of the constructed range profiles are tested by employing the same CFAR detector used for target detection in Fig.~\ref{fig:exp_RP_detection}\subref{fig:exp_RP_detec_BefIM}. Fig.~\ref{fig:exp_RP_detection}\subref{fig:exp_RP_detec_AftIM_wd}-\subref{fig:exp_RP_detec_AftIM_CFAR_AC} show the output results of the CFAR detector. One can see that the three peaks related to the aggressor radar and two TCRs are all detected after IM with all the four methods while the TCR at the further distance was missed based on the range profile before IM (Fig.~\ref{fig:exp_RP_detection}\subref{fig:exp_RP_detec_BefIM}). So the four IM methods improve the targets' probability of detection. Moreover, based on the range profiles obtained with the CFAR-Z and CFAR-AC, a fourth target, which is a stationary car at a distance of $22.5\,\mathrm{m}$, is also detected (Fig.~\ref{fig:exp_RP_detection}\subref{fig:exp_RP_detec_AftIM_CFAR_Z} and \subref{fig:exp_RP_detec_AftIM_CFAR_AC}) but missed when the RPs acquired with the WD and ANC methods were used (Fig.~\ref{fig:exp_RP_detection}\subref{fig:exp_RP_detec_AftIM_wd} and \subref{fig:exp_RP_detec_AftIM_ANC}). Therefore, the beat signals obtained with the CFAR-Z and CFAR-AC IM approaches provide higher target's probability of detection than those recovered with the WD and ANC methods.

\section{Conclusion} \label{sec:conclusion}

In the paper, we proposed two CFAR-based approaches, i.e., CFAR-Z and CFAR-AC, to mitigate inference for FMCW radars system, which exploit the CFAR detector to detect the large chirp-pulse like interferences in the time-frequency domain and then apply the zeroing and amplitude correction for mitigate them, respectively. Compared to the prior art methods, both approaches achieve better interference mitigation performance in terms of both SINR and correlation coefficient of the recovered signal after IM. Moreover, both CFAR-Z and CFAR-AC approaches are computationally efficient and could be implemented for real-time processing for automotive radars. 


%



\section*{Acknowledgment}

The authors would like to thank Ms Y. Lu and Mr I. R. Montero for their help during the experimental measurement.




\bibliographystyle{IEEEtran}



%

\bibliography{Reference}

%








\end{document}